\documentclass[aps,pra,showpacs,amssymb,nofootinbib,twocolumn,longbibliography]{revtex4-1}
\pdfoutput=1
\usepackage[utf8]{inputenc}
\usepackage[english]{babel}

\usepackage{tikz}
\usetikzlibrary{shapes,backgrounds,fit,decorations.pathreplacing,arrows,decorations.markings}
\usepackage{verbatim}
\tikzstyle{vecArrow} = [thick, decoration={markings,mark=at position
   1 with {\arrow[semithick]{open triangle 60}}},
   double distance=1.4pt, shorten >= 5.5pt,
   preaction = {decorate},
   postaction = {draw,line width=1.4pt, white,shorten >= 4.5pt}]
\tikzstyle{innerWhite} = [semithick, white,line width=1.4pt, shorten >= 4.5pt]
\usepackage{bbm}
\usepackage{bm}
\usepackage{amsbsy}
\usepackage{amsthm}
\usepackage{amssymb}
\usepackage{amsfonts}
\usepackage{amsmath}
\usepackage{dsfont} 
\usepackage{graphicx} 
\usepackage{epsfig}
\usepackage{epstopdf}
\usepackage{dsfont}
\usepackage{multibib}
\usepackage{color}

\usepackage[colorlinks]{hyperref}
\usepackage{xcolor}
\usepackage{soul}
\makeatletter
\newcommand\org@hypertarget{}
\let\org@hypertarget\hypertarget
\renewcommand\hypertarget[2]{%
  \Hy@raisedlink{\org@hypertarget{#1}{}}#2%
  }
\makeatother
\usepackage[figure,table]{hypcap}
\usepackage{MnSymbol}
\usepackage{enumerate}
\usepackage{float}
\hypersetup{
	bookmarksnumbered,
	pdfstartview={FitH},
	citecolor={darkgreen},
	linkcolor={darkred},
	urlcolor={darkblue},
	pdfpagemode={UseOutlines}}
\definecolor{darkgreen}{RGB}{50,190,50}
\definecolor{darkblue}{RGB}{0,0,190}
\definecolor{darkred}{RGB}{238,0,0}
\definecolor{quantum}{RGB}{83,37,127}
\definecolor{quantumlight}{RGB}{169,146,191}
\usepackage{soul}

\newcommand{\ket}[1]{\ensuremath{\left|\right.\!{#1}\!\left.\right\rangle}}

\newcommand{\bra}[1]{\ensuremath{\left\langle\right.\!{#1}\!\left.\right|}}

\newcommand{\ketbra}[2]{\ensuremath{|{#1}\rangle\!\langle{#2}|}}

\newcommand{\subtiny}[3]{\ensuremath{_{\hspace{#1 pt}\protect\raisebox{#2 pt}{\tiny{$ #3$}}}}}

\newcommand{\tr}{\textnormal{Tr}}

\newcommand{\djj}{d\kern-0.4em\char"16\kern-0.1em}
\renewcommand{\thesection}{\Roman{section}}
\renewcommand{\thesubsection}{\Roman{section}.\arabic{subsection}}
\renewcommand{\thesubsubsection}{\Roman{section}.\arabic{subsection}.\arabic{subsubsection}}
\makeatletter
\renewcommand{\p@subsection}{}
\renewcommand{\p@subsubsection}{}
\makeatother

\renewcommand{\hl}[1]{#1}

\usepackage[customcolors]{hf-tikz}
\tikzset{style green/.style={
    set fill color=green!50!lime!60,
    set border color=white,
  },
  style cyan/.style={
    set fill color=cyan!90!blue!60,
    set border color=white,
  },
  style orange/.style={
    set fill color=orange!80!red!60,
    set border color=white,
  },
  style hordash/.style={
    set fill color=white,
    set border color=black,
  },
     style rose/.style={
    set fill color= magenta!70!pink!70, 
    set border color=white,
  },
  hor/.style={
    above left offset={-0.09,0.25},
    below right offset={0.09,-0.05},
    #1
  },
  ver/.style={
    above left offset={-0.09,0.35},
    below right offset={0.09,-0.1},
    #1
  }
}

\usepackage[framemethod=tikz]{mdframed}
\definecolor{mycolor}{rgb}{0.122, 0.435, 0.698}
\newmdenv[innerlinewidth=0.5pt, roundcorner=4pt,linecolor=mycolor,innerleftmargin=6pt,
innerrightmargin=6pt,innertopmargin=6pt,innerbottommargin=6pt]{mybox}
\usepackage{paracol}
\usepackage{tcolorbox}
\tcbuselibrary{theorems}

\newtcbtheorem{Definitions}{Definition}%
{colback=mycolor!5,colframe=mycolor,fonttitle=\bfseries,
left=4pt,
right=4pt,
top=5pt,
bottom=5pt}{defi}

\newtcbtheorem{Lemmas}{Lemma}%
{colback=green!5,colframe=green!50!blue,fonttitle=\bfseries,
left=4pt,
right=4pt,
top=5pt,
bottom=5pt
}{lemma}
\newtcbtheorem{Results}{Result}%
{colback=orange!5,colframe=orange!50!blue,fonttitle=\bfseries,
left=4pt,
right=4pt,
top=5pt,
bottom=5pt,
title={#2},
}{result}
\newtcolorbox[blend into=figures]{boxdefi}[3][]
{ float*=ht,width=\textwidth,lower separated=false, center upper,
title={#2},label= def:#3,#1}

\begin{document}

\title{Autonomous Temporal Probability Concentration:\\
Clockworks and the Second Law of Thermodynamics}
\author{Emanuel Schwarzhans}
\email{emanuel.schwarzhans@oeaw.ac.at}
\affiliation{Institute for Quantum Optics and Quantum Information - IQOQI Vienna, Austrian Academy of Sciences, Boltzmanngasse 3, 1090 Vienna, Austria}
\author{Maximilian P. E. Lock}
\affiliation{Institute for Quantum Optics and Quantum Information - IQOQI Vienna, Austrian Academy of Sciences, Boltzmanngasse 3, 1090 Vienna, Austria}
\author{Paul Erker}
\affiliation{Institute for Quantum Optics and Quantum Information - IQOQI Vienna, Austrian Academy of Sciences, Boltzmanngasse 3, 1090 Vienna, Austria}
\author{Nicolai Friis}
\affiliation{Institute for Quantum Optics and Quantum Information - IQOQI Vienna, Austrian Academy of Sciences, Boltzmanngasse 3, 1090 Vienna, Austria}
\author{Marcus Huber}
\email{marcus.huber@univie.ac.at}
\affiliation{Institute for Quantum Optics and Quantum Information - IQOQI Vienna, Austrian Academy of Sciences, Boltzmanngasse 3, 1090 Vienna, Austria}
\affiliation{Vienna Center for Quantum Science and Technology, Atominstitut, TU Wien, 1020 Vienna, Austria}


\begin{abstract}
According to thermodynamics, the inevitable increase of entropy allows the past to be distinguished from the future. From this perspective, any clock must incorporate an irreversible process that allows this flow of entropy to be tracked. In addition, an integral part of a clock is a clockwork, that is, a system whose purpose is to temporally concentrate the irreversible events that drive this entropic flow, thereby increasing the accuracy of the resulting clock ticks compared to counting purely random equilibration events. In this article, we formalise the task of \emph{autonomous temporal probability concentration} as the inherent goal of any clockwork based on thermal gradients. Within this framework, we show that a perfect clockwork can be approximated arbitrarily well by increasing its complexity. Furthermore, we combine such an idealised clockwork model, comprised of many qubits, with an irreversible decay mechanism to showcase the ultimate thermodynamic limits to the measurement of time.
\end{abstract}

\maketitle

\section*{Popular summary}\vspace*{-2mm}
\textbf{ The laws of physics governing the microscopic world from which our experience emerges know no directionality of time. A recording of an isolated physical system looks plausible played forwards and backwards. Nevertheless, an arrow of time can be identified in many complex processes, such as the breaking of an egg. The irreversibility of such processes can also be used to keep track of the passage of time. In our paper, we investigate the implications of this insight for building (quantum) clocks from first principles. Any system that can function as a clock needs an irreversible component whose dynamics indicate a clear directionality of time. However, typical irreversible processes found in nature, such as decaying atoms and equilibrating heat baths, have no discernable regular temporal structure and therefore make for rather bad clocks by themselves. It thus comes at no surprise that cooling coffee cups and similar items are not usually used for timekeeping. Instead, one relies on the combination of internal, periodic dynamics with an irreversible, temporally unstructured, process to construct well-functioning clocks. The same is fundamentally true when envisioning clocks at the quantum scale: periodic internal dynamics of a clockwork are combined with an irreversible process to yield temporally well-structured ‘ticks’ of a clock. In this work, we identify and systematically analyse the task performed by the clockwork, which we refer to as autonomous temporal probability concentration (ATPC). That is, the clockwork generates a periodic structure that determines at which times the irreversible events -- the clock’s ticks -- can occur. We show that ATPC is intimately related to the complexity of the clockwork and can be performed arbitrarily well, as long as complexity is not constrained. Nonetheless, the irreversible process itself sets a fundamental limit for the quality of clocks, beyond which they cannot be improved. In this sense, the second law of thermodynamics -- the origin of irreversibility, generates but also limits the potential for any system to serve as a clock, while the complexity of the clockwork determines how well this can be achieved in practice.
} 
\section{Introduction}\label{sec:introduction}\vspace*{-2mm}

Time plays a special role in quantum physics. While other physical quantities of interest are represented as Hermitian operators, there is no observable corresponding to time itself. That is, it is not possible to find an operator conjugate to the Hamiltonian (representing energy) that may serve as `time observable' in the same way as is done for position and momentum~\cite{PauliStraumann1990} (see e.g.~\cite{garrison1970canonically} for some caveats to this statement). Time thus plays the role of a parameter in the equations of motion.{ Consequently, the passage of time is estimated via the evolution of a reference system} \textemdash\ a \emph{clock}. By tracking the dynamical evolution of (observable quantities related to) such a clock system it is possible to extract information about the flow of time, see, \hl{e.g.~}\cite{braunstein1996generalized,milburn2017quantum,Erkerthesis,PhysRevX.6.041053,khandelwal2020universal,alex1,alex2,milburn2020thermodynamics}. \emph{But what makes a specific system useful as a clock?}
\begin{figure*}[tbp]
  \centering
  \includegraphics[width=\textwidth]{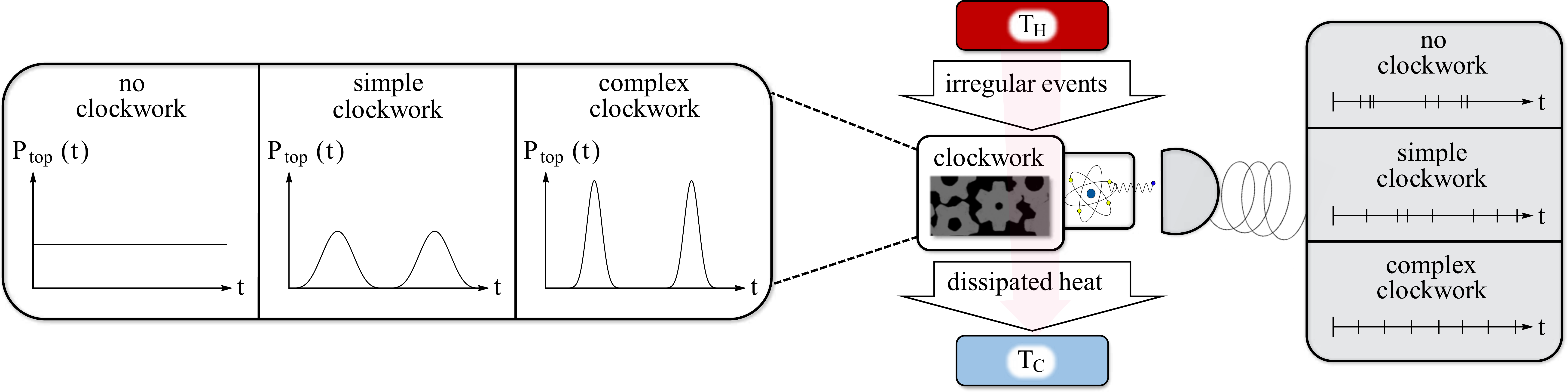}
  \vspace*{-2mm}
  \caption{{\footnotesize Illustration of timekeeping at the level of individual irreversible events. The equilibration events that follow the second law of thermodynamics are inherently stochastic and irregular, in our example we use radiative decays from an excited energy level of a quantum system (which we label `top'). By inserting an autonomously operating clockwork between the two out-of-equilibrium systems ('hot' decaying quantum systems and a 'cold' environment), these decays are temporally structured by temporal variation of the population of the top level, a task that we refer to as Autonomous Temporal Probability Concentration (ATPC). This concentrates the probability of such a decay around the oscillatory peaks of excited population. The panel on the right showcases how greater clockwork complexity leads to a regularisation of individual thermalisation events, \emph{i.e.}, clock ticks. Starting from a thermal population with randomly distributed ticks depicted above, continuing to a simple clockwork with limited population and still significant variance, resulting in ticks being more likely during peak populations and thus less frequent and more regular, and then finally a complex clockwork, increasing population while decreasing temporal variance, giving yet more regular ticks. The tick distribution on the right is exemplary and depicts random ticks, whose spacing approximates the cycle time of the clock work as the temporal probability becomes more concentrated around sharp peaks}}
  \label{fig:cartoon}
\end{figure*}
To address this question, we consider time to be a continuously elapsing parameter $t$ (`Schr{\"o}dinger time') whose value is estimated by a clock in terms of discrete increments (`ticks'). According to quantum theory, the evolution of any closed system is time-reversal symmetric, and therefore any complete description of an instrument that measures time inevitably requires an irreversible part that breaks this symmetry. By definition, the equilibrium state of any system features no non-trivial evolution in time. Thus, the first necessary ingredient for building a clock is an out-of-equilibrium system, such that the clock can harness the irreversible transition to higher entropy to produce ticks.

Entropy-increasing processes are fundamentally stochastic. Consequently, individual events resulting from such a process provide little information about $t$ and thus make for rather bad clocks. While one could, in principle, use any equilibrating system as a clock \textemdash\  such as a hot coffee mug cooling down on your desk \textemdash\ its ticks, e.g.\hl{,} the spontaneous emissions of thermal photons (which exhibit super-Poissoninan statistics), come at highly irregular intervals with respect to Schr{\"o}dinger time. Structuring this irregular entropy flow into a series of ticks to allow for a precise synchronisation of events is exactly the purpose of a clock. In this article we formalise the task of timekeeping by conceptually separating two stages:

\begin{itemize}
    \item[(i)] an irreversible process that follows the second law of thermodynamics, \emph{i.e.} an out-of-equilibrium system moving towards equilibrium by means of discrete and stochastic events,
    \item[(ii)] an internal clockwork that temporally concentrates the probability of an irreversible event occurring, thereby \hl{mitigating the fluctuations of the intervals between the equilibration events.}
\end{itemize}

As we will see, the particular choice of (i) provides the context for evaluating clock performance because it represents a basic form of clock itself, while at the same time limiting the performance of a clock for any given clockwork. Stage (ii) gives rise to a clearly defined mathematical task that we will refer to as \emph{autonomous temporal probability concentration} (ATPC).

\hl{Here, we consider clocks to be autonomous. That is, the Hamiltonian generating the evolution of the clockwork is energy-conserving and time-independent, and the irreversible process is memoryless and requires no external control, \emph{i.e.}, \emph{no active measurement}. Although current quantum clocks are usually far from autonomous, as they require power input and are subject to losses, both of which are usually not fully accounted for in their analysis, we focus on autonomous clocks in order to provide a full analysis of the resources that are fundamentally required to drive a clock.} 

\hl{To 
describe} the performance of a clock, we use two quantities: \emph{accuracy} and \emph{resolution}. The accuracy \hl{${N=\left(\tfrac{\bar{t}}{\Delta t}\right)^2}$}, is the \hl{average} number of ticks until the clock is, on average, off by one tick with respect to Schr{\"o}dinger time. The resolution ${R=1/\bar{t}}$, is the average of the tick frequency with respect to Schr{\"o}dinger time. 
\hl{Note that this choice for quantifying the resolution is not the only possibility. We choose the above definition since it represents a conservative figure of merit in the sense that it prevents statistical outliers of ticks occurring at small times $t$ from unduly inflating the estimated resolution, as we discuss in more detail in Sec.}~\ref{sec:Irreversibility}.

That there is a trade-off relation between accuracy and resolution, and that there is a proportional relation between the entropy dissipated in the process and the clock performance, was first noticed in a model of an autonomous quantum clock as an open quantum system in~\cite{ErkerMitchisonSilvaWoodsBrunnerHuber2017} and recently corroborated in a mesoscopic experiment in~\cite{PearsonEtal2020}.

Here, we combine these aspects and provide a detailed investigation of the trade-offs between accuracy, resolution, and entropy production for given energy and complexity within the framework of autonomous quantum clocks~\cite{Goold_2016,Mitchison_2019}.
A central tenet for providing these trade-offs is the separation of timekeeping into two separate processes mentioned above: (i) the irreversible out-of-equilibrium transitions of the clockwork via interaction with an environment, resulting in distinguishable events registered as `ticks', which we model with a decay mechanism, and (ii) the internal closed-system (unitary) dynamics that provide a clockwork and temporally concentrate the population of states from which an irreversible transition can emerge. That is, the clockwork ensures that the circumstances that allow for a tick to happen (\emph{e.g.}, a specific energy level resonant with the out-of-equilibrium dynamics being highly populated) occur only within a very narrow time window. 

We first find that a simple clockwork can only concentrate probability in a limited fashion, prompting the question of whether more complex designs could perform better. We answer this by finding an analytical relation between ATPC and complexity of a specific clockwork model. More generally, we identify the important features of a clockwork that lead to this improvement, and prove that for cold environments, ATPC can be performed arbitrarily well. Then we investigate whether perfect ATPC allows for perfect clocks and find that the answer is no. In fact, the irreversible process sets a limit to the clock quality, and while increasing the complexity (and thus the concentration of probability) first improves the quality of the clock, after a certain point a further increase will actually be detrimental. We thus illustrate the trade-offs between accuracy, resolution, entropy production, and clockwork complexity.

The specific clock model that we consider here consists of (1) external heat baths as out-of-equilibrium resources, (2) a quantum system representing the `clockwork', and (3) an external field that the clockwork can emit energy (`ticks', \emph{e.g.}, photons) into. In Sec.~\ref{sec:framework}, we first discuss the role and choice of the clockwork, and formalise the task of ATPC. In Sec.~\ref{sec:Irreversibility}, we then discuss mechanisms for coupling the clockwork to an equilibrating process to produce ticks. In Sec.~\ref{sec:Numerical results} we combine the two, to showcase the limitations set by the irreversible process and how the complexity of a clockwork can be utilised to reach the maximal potential of a clock. We continue in Sec.~\ref{sec:discussion} with a discussion of the implications and the relation to other literature on clocks and end with a short conclusion in Sec.~\ref{sec:conc}.


\section{Thermal machines and the clockwork}\label{sec:framework}


Let us now consider a clockwork in the sense discussed above, that is, a device that contains a target subsystem, which is to be prepared for an out-of-equilibrium transition, thus resulting in a `tick'. From a thermodynamic perspective, such a preparation requires work to be performed on the target, which can be achieved by a quantum thermal machine. Operating such a machine in turn requires an out-of-equilibrium resource, which we here consider to be provided by thermal baths at different temperatures, \emph{i.e.} a thermal gradient. More specifically, we assume that two independent baths are available, a hot bath and a cold bath, at temperatures $T_{\mathrm{H}}$ and $T_{\mathrm{C}}$, respectively, where the latter represents the environment. This setup is depicted in Fig.~\ref{fig:cartoon}.

This choice is motivated, first, by the general availability of heat baths, \hl{i.e. it is the most common out-of-equilibrium resource found in nature, such as e.g. the sun}. Second, because systems are usually expected to thermalise (eventually) without \hl{detailed external control or timing}, i.e. preparing such heat baths does not require any timing device or detailed control of the system's internal structure, just an increase in average energy. Consequently, heat baths allow for transparent bookkeeping of the relevant resources, \emph{i.e.} of the average amount of entropy dissipated by the clockwork for each tick. 

A specific focus of the analysis performed here lies on the identification of trade-offs between different figures of merit for the clock performance for fixed energy input and clock complexity. In principle, the performance of a given clock also depends on the (difference between the) temperatures $T_{\mathrm{C}}$ and $T_{\mathrm{H}}$. However, since we are primarily interested in upper bounds on the relevant figures of merit, we will often concentrate on the case where the environment temperature is $T_{\mathrm{C}}=0$. For the sake of completeness, calculations for general $T_{\mathrm{C}}$ can be found in Appendices~\ref{appsec: The horizontal extension} and~\ref{appsec:vertical extension details}.

Our clockwork model then consists of two parts, a $d$-dimensional `\emph{ladder}' target system (in the simplest case, a qubit, $d=2$) and a machine, which itself has some substructure and couples to the ladder via unitary interaction. 
This interaction supplies work (which the machine draws from its coupling to the heat baths) to the ladder, driving it to its excited states. The ladder in turn couples \hl{irreversibly} to an external field, and thus these excitations eventually result in ticks (\emph{i.e.} energy emitted into the field). Here, we consider a model where only a non-zero population $P_{\mathrm{top}}(t)$ of the `top level' \textemdash\ the most highly excited state of the ladder \textemdash\ can lead to a tick. \hl{Barring some improbable combination of selection rules, such a single sharp energy transition can in practice of course only be approximated. However, as will become clear once we have introduced our model, allowing the possibility of clock ticks occurring due to other transitions would serve to spread the temporal profile of the ticks, decreasing probability concentration. In the spirit of deriving idealised but fundamental bounds, we therefore focus on decays resulting from only one particular transition.
} As a consequence, the quality of the clockwork depends on the properties of the particular probability distribution $P_{\mathrm{top}}(t)$ as a function of Schr{\"o}dinger time $t$. In particular, an ideal clockwork should be capable of producing
\begin{align}
    P_{\mathrm{top}}(t)=
    \begin{cases}
        1,& \text{if } t=t_0\\
        0,              & \text{otherwise}
    \end{cases}.
    \label{eq:ideal ptop}
\end{align}
While one would expect a perfect clockwork to be capable of producing this distribution, it is also clear that it is not always desirable in conjunction with an irreversible mechanism. \hl{If the probability is arbitrarily temporally concentrated, \emph{i.e.}, it is only} \hl{close to one for a short period of time}\hl{, but the coupling of the ladder to the external field is of finite strength, then the emission of the ladder energy into the field has a chance not to occur during the peak, thus skipping this tick and worsening the clock performance.} Nonetheless, an ideal clockwork should be capable of approximating this ideal distribution to the desired precision set by the irreversible mechanism. 
Arguably, it seems implausible that a heat engine itself, which intrinsically also harnesses the stochastic flow of energy from a hot to a cold bath, should be able to produce such a perfect signal. However, it may be reasonable to expect that a sufficiently complex clockwork, itself driven by a heat engine, could approximate the ideal ATPC of Eq.~(\ref{eq:ideal ptop}). In the following, we therefore investigate the role of the complexity of the internal structure of the machine in approximating the ideal ATPC. In order to do so, we decompose the machine into a set of elementary few-qubit machines, each realising an effective virtual qubit~\cite{BrunnerLindenPopescuSkrzypczyk2012}. This allows the number of (elementary) machines to be used as a proxy for the complexity of the clockwork's microscopic structure. In terms of these quantifiers, \emph{i.e.} the dimension $d$ of the target system and the number $M(d-1)$ of virtual-qubit machines\footnote{We consider each of the $d-1$ transitions between neighbouring energy levels of the ladder to be coupled to $M$ virtual-qubit machines.}, a central result on autonomous probability concentration that we derive in this paper can be phrased as follows:\\

\begin{Results}{Autonomous temporal probability concentration of qubit machines}{1}
Driving a $d$-dimensional target system at temperature $T_\mathrm{C}=0$, with $M$ virtual-qubit machines per transition between neighbouring levels, autonomously allows a top-level probability of 
\begin{align} \label{eq:PtopGen}
    P_{\mathrm{top}}(t) &=\,\Bigl(1-\Bigl(1-\Bigl(\frac{\mathcal{Z}_{\mathrm{H}}-1}{\mathcal{Z}_{\mathrm{H}}}\Bigr)^{d-1}\Bigr)^{M}\Bigr) \sin^{2(d-1)}(gt)
\end{align}
to be reached. Here, $\mathcal{Z}_{\mathrm{H}}$ is the partition function of a qubit coupled to the hot bath, and can thus take values between $1$ and $2$.
\end{Results}

In other words, we show in the following that the behaviour of an ideal clockwork [\emph{i.e.} Eq.~(\ref{eq:ideal ptop})] can be approximated arbitrarily well by increasing the complexity of the clockwork, that is, by increasing~$M$ and~$d$.


\subsection{Two-Qubit Machine}\label{Two-Qubit Machine}

\begin{figure}[ht!]
  \centering
  \includegraphics[width=0.4\textwidth]{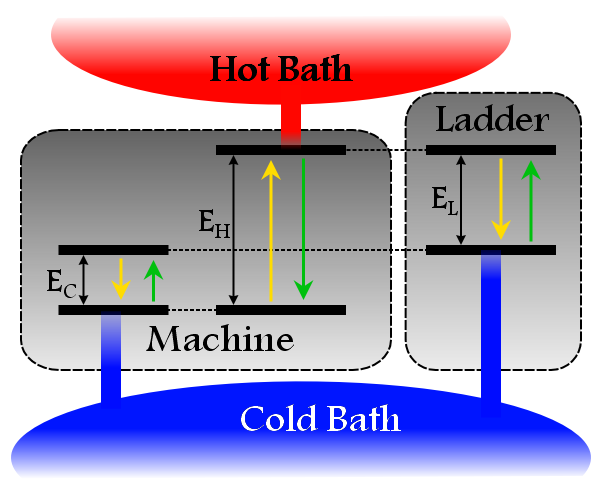}
  \vspace*{-1.5mm}
  \caption{{\footnotesize Energy-level structure of the minimal thermal clockwork. The transitions induced by $H_{\mathrm{int}}$ are indicated by arrows. The green arrows indicate the transition where the ladder gets exited. The yellow arrows show the reverse transition. Coupling the qubit with the biggest energy gap to the hot bath introduces a bias towards the transition that is indicated by the green arrows.}}
  \label{fig:Minimalmachine_2}
\end{figure}

We begin by considering the simplest possible heat-engine-driven clockwork: a $2$-dimensional ladder coupled to a `cold' bath (the environment) and to a two-qubit machine, \emph{i.e.} $d=2$ and $M=1$. In terms of Hilbert space dimension, this is the smallest possible thermal machine~\cite{BrunnerLindenPopescuSkrzypczyk2012}, consisting of a `cold' qubit and a `hot' qubit, in contact with the `cold' environment and a hot bath, respectively, as illustrated in Fig.~\ref{fig:Minimalmachine_2}.

Before the machine is activated, the qubits only interact with their respective baths. Under the assumption of weak coupling between the qubits and baths, each qubit thermalises to the corresponding bath temperature. Denoting the energy gaps of the hot, cold and ladder qubits as $E_{\mathrm{H}}$, $E_{\mathrm{C}}$ and $E_{\mathrm{L}}$ respectively, the reduced states of the qubits can be represented by the thermal states
\begin{align}
    \rho_{i}=\frac{e^{-\beta_{i} H_{i}}}{\mathcal{Z}_{i}},
\end{align}
with $i=\mathrm{H,C,L}$, and where $\mathcal{Z}_{i}=1+e^{-\beta_{i}E_{i}}$ are the respective partition functions and $H_{i}$ the corresponding free Hamiltonians with eigenstates $\ket{0_{i}}$ and $\ket{1_{i}}$. The total initial state of the clockwork \textemdash\ the machine and the ladder \textemdash\ 
thus takes the form $\rho_0=\rho_{\mathrm{H}}\otimes\rho_{\mathrm{C}}\otimes\rho_{\mathrm{L}}$.
\begin{figure*}[tbp]
  \centering
  \includegraphics[width=0.97\textwidth]{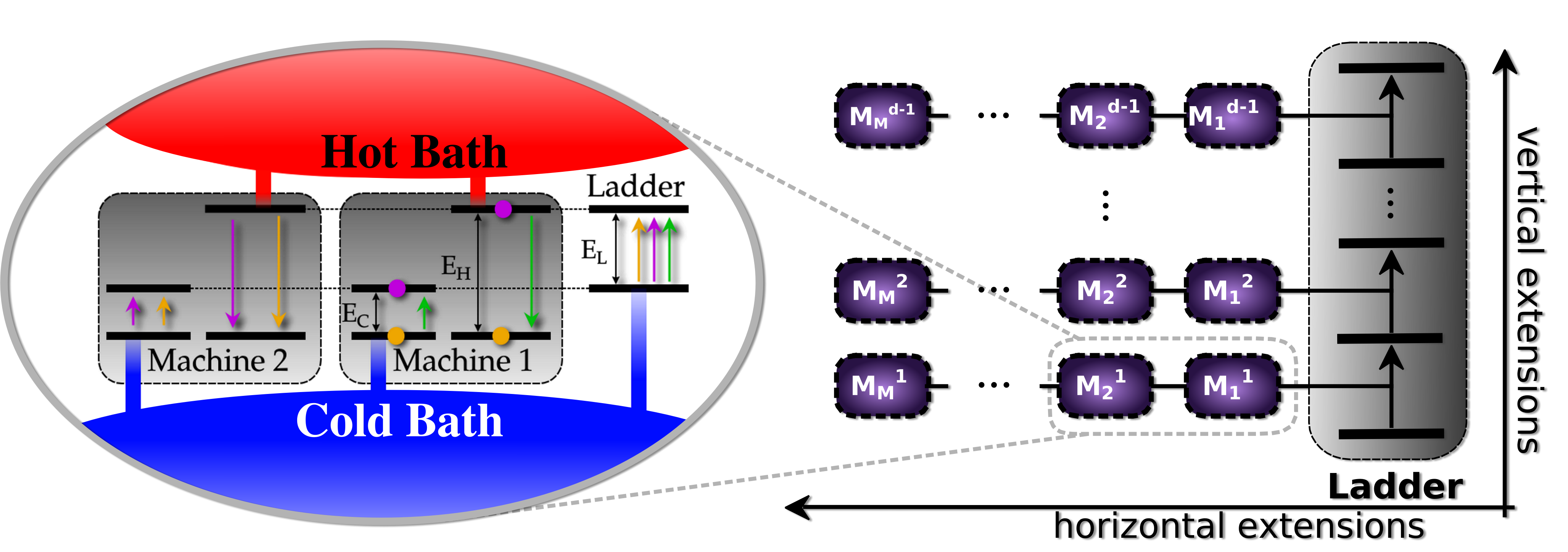}
  \vspace*{-2mm}
  \caption{{\footnotesize 
  On the right-hand side, the notion of `horizontal' and `vertical' extension is schematically illustrated. The ladder dimension $d$ determines the number of vertical extensions (`rows'). Each pair of neighbouring energy levels of the ladder couples to $M$ copies of the two-qubit machine, where $M$ determines the number of horizontal extensions (`columns'). 
  On the left-hand side, the energy-level structure of a horizontally extended machine for a two-machine clockwork coupling to a two-level ladder ($d=2$) is illustrated. This is a special case of the generalised clockwork shown on the right-hand side. The different energy-conserving transitions induced by $H_{\mathrm{int}}$ are indicated by the differently coloured arrows and dots, where the dots represent a conditioning of the transitions of machine 2 either on the ground state (orange) or on the exited state (purple) of machine 1. The reverse transitions are not depicted here for sake of clarity. 
  }}
  \label{horvert}
\end{figure*}

We further assume that the timescale of the interaction between the machine and target qubits is much shorter than that of their thermalisation with the respective baths. Consequently, the relevant dynamics of the clockwork are well described by energy-conserving unitary processes on the clockwork Hilbert space $\mathcal{H}=\mathcal{H}_{\mathrm{H}}\otimes\mathcal{H}_\mathrm{C}\otimes\mathcal{H}_{\mathrm{L}}$. \hl{ This corresponds to assuming that the energy scales of the clockwork are much greater than that of its coupling to the environment. It is suited to our purpose of obtaining fundamental limits to the task of ATPC, as relaxing this assumption will result in the regularity of the clockwork being impaired by the randomness of the bath. Now, since the purpose of the machine is to transfer energy to the target system, we are} interested in designing the internal structure of the clockwork, namely the energy levels of the free Hamiltonian $H_{0}=H_{\mathrm{H}}+H_{\mathrm{C}}+H_{\mathrm{L}}$ and an interaction Hamiltonian $H_{\mathrm{int}}$ such that $\left[H_{0},H_{\mathrm{int}}\right]=0$ and $\left[H_{\mathrm{L}},H_{\mathrm{int}}\right]\neq0$. This can be achieved by choosing the energy gaps to satisfy $E_{\mathrm{H}}\geq E_{\mathrm{C}}$ and $E_{\mathrm{L}}=E_{\mathrm{H}}-E_{\mathrm{C}}$, which results in two degenerate energy levels of $H_{0}$: $\ket{0_{\mathrm{C}}1_{\mathrm{H}}0_{\mathrm{L}}}$ and $\ket{1_{\mathrm{C}}0_{\mathrm{H}}1_{\mathrm{L}}}$. This, in turn, allows us to define an interaction Hamiltonian that acts non-trivially only within the degenerate subspace, given by
\begin{align}
    H_{\mathrm{int}}    &=g\bigl(\ket{1_{\mathrm{C}}0_{\mathrm{H}}1_{\mathrm{L}}}\!\!\bra{0_{\mathrm{C}}1_{\mathrm{H}}0_{\mathrm{L}}}+\ket{0_{\mathrm{C}}1_{\mathrm{H}}0_{\mathrm{L}}}\!\!\bra{1_{\mathrm{C}}0_{\mathrm{H}}1_{\mathrm{L}}}\bigr), \label{eq:minimal_hint}
\end{align}
where $g \in \mathds{R}$ is a coupling constant. The unitary dynamics generated by the total Hamiltonian $H=H_{0}+H_{\mathrm{int}}$ hence conserves the total energy of the clockwork, since $\left[H_{0},H_{\mathrm{int}}\right]=0$. However, since $\left[H_{\mathrm{L}},H_{\mathrm{int}}\right]\neq0$, the interaction, once activated, can perform work on the ladder. 

The resulting dynamics leads to an increase of the population of the top energy level $\ket{1_{\mathrm{L}}}$ of the ladder, which (in units where $\hbar=1$) is given by 
\begin{align}
    P_{\mathrm{top}}(t) &=
    \tr\bigl(\ket{1_{\mathrm{L}}}\!\!\bra{1_{\mathrm{L}}} e^{-iHt}\rho_{0}\,e^{iHt}\bigr).
    \label{eq:def top level prob}
\end{align}
The maximally reachable population depends on the temperatures of the baths, as well as the energy gaps of the machine qubits~\cite{BrunnerLindenPopescuSkrzypczyk2012}. The top-level probability in Eq.~(\ref{eq:def top level prob}) evaluates to (see Appendix~\ref{appesec:top-level probability of a two-qubit clockwork})
\begin{align}
    P_{\mathrm{top}}(t) &=\Bigl(\tfrac{\mathcal{Z}_{\mathrm{H}}-1}{\mathcal{Z}_{\mathrm{C}}\mathcal{Z}_{\mathrm{H}}\mathcal{Z}_{\mathrm{L}}}\Bigl)\,\sin^2(g t)
    +\Bigl(\tfrac{(\mathcal{Z}_{\mathrm{C}}-1)(\mathcal{Z}_{\mathrm{L}}-1)}{\mathcal{Z}_{\mathrm{C}}\mathcal{Z}_{\mathrm{H}}\mathcal{Z}_{\mathrm{L}}}\Bigl)\,\cos^2(g t)
    \nonumber\\
    &\ \ +\tfrac{\mathcal{Z}_{\mathrm{L}}-1}{\mathcal{Z}_{\mathrm{L}}}-\tfrac{(\mathcal{Z}_{\mathrm{C}}-1)(\mathcal{Z}_{\mathrm{L}}-1)}{\mathcal{Z}_{\mathrm{C}}\mathcal{Z}_{\mathrm{H}}\mathcal{Z}_{\mathrm{L}}}\,.
    \label{eq:ptop 2qb machine general}
\end{align}
For $T_{\mathrm{C}}=0$, this simplifies to
\begin{align}
    P_{\mathrm{top}}(t) =\Bigl(1-\frac{1}{\mathcal{Z}_{\mathrm{H}}}\Bigl)\,\sin^2(g t)\,.
    \label{eq:ptop 2qb machine}
\end{align}

Thus, even when $T_\mathrm{C}=0$, this function is far away from the ideal shape in Eq.~(\ref{eq:ideal ptop}), both in terms of its maximal value and the width of the distribution around its peak. Even in the limit $T_{\mathrm{H}}\rightarrow \infty$, the maximal value reached at $t=\frac{\pi}{2g}$ is only $\tfrac{1}{2}$. Moreover, this top-level population could also have been achieved by directly coupling the ladder to the hot bath. Thus, the two-qubit machine does not provide the desired ATPC by itself. However, in the following we present a generalisation of this framework which allow arbitrarily precise ATPC, and hence an ideal clockwork to be approximated to within any given error.\\


\vspace*{-2mm}
\subsection{Generalised Machines}\label{sec:Generalised Machines}
\vspace*{-2mm}

In the following we present a generalised clockwork model that allows both the `sharpness' and the amplitude of $P_{\mathrm{top}}(t)$ to be controlled, while keeping track of all the relevant resources. This can be achieved by two qualitatively different but compatible extensions that we refer to as `horizontal' and `vertical' extensions, as illustrated in Fig.~\ref{horvert}. The horizontal extension allows the amplitude of $P_{\mathrm{top}}(t)$ to be increased, while the vertical extension allows the width of the peak of $P_{\mathrm{top}}(t)$ to be decreased, thus increasing its `sharpness'. Specifically, we add more levels to the target ladder and with it more two-qubit machines, interacting with each successive transition (vertical extension); to a given transition we add more machines (horizontal extension). We start by collecting all interactions along a vertical column (see Fig.~\ref{horvert}) of machines interacting with the ladder into a term $H_1$. This vertically extends the interaction of a single two-qubit machine, Eq.~\eqref{eq:minimal_hint}, along all ladder states, \emph{i.e.}\begin{align}
    H_{1}    
    &=g\sum_{n=1}^{d-1}\bigl(\ket{1_{\mathrm{C}}0_{\mathrm{H}}}\!\!\bra{0_{\mathrm{C}}1_{\mathrm{H}}}_{\mathrm{M}_1^n}\otimes \ket{n+1_{\mathrm{L}}}\!\!\bra{n_{\mathrm{L}}}+\text{H.c.}\bigr),
\end{align}
for the first vertical column, where $\mathrm{M}_i^j$ denotes the Hilbert space of the $j^\mathrm{th}$ two-qubit machine acting on the $i^\mathrm{th}$ ladder transition. We then add another term $H_2$, which does the same for the second vertical column, albeit with an additional projector \hl{onto the subspace orthogonal to the one on which $H_1$ acts non-trivially} to ensure commutativity of $H_1$ and $H_2$. This continues for $M$ vertical columns, always projecting onto the orthogonal subspace of all previously \hl{added} machines. Using M$_{(i)}$ to denote the Hilbert space of the vertical group of the $i
^\mathrm{th}$ machine, \emph{i.e.} $\mathrm{M}_{(i)}:=\bigotimes_{j=1}^{d-1}\mathrm{M}_i^j$, we can then write our generalisation of the interaction Hamiltonian from the previous section in a compact notation as
\begin{align}\label{generalHamiltonian}
    H_{\mathrm{int}}=\sum_{k=1}^{M} \; 
    \bigotimes_{i=1}^{k-1}\mathbbm{1}_{\mathrm{M}_{(i)}}\otimes J_{\mathrm{M}_{(k)}\mathrm{L}} \otimes \bigotimes_{i=k+1}^M \Pi_{\mathrm{M}_{(i)}} 
    \,=\,
    \sum_{k=1}^{M} H_k\,.
\end{align}
Here, we have defined the projectors
\begin{align}
{\Pi_{\mathrm{M}_{(i)}} := \mathbbm{1}_{\mathrm{M}_{(i)}}-\sum_{n=0}^{d-1} \ketbra{n_{\mathrm{M}_{(i)}}}{n_{\mathrm{M}_{(i)}}}}
\end{align}

and the operator
\begin{align}
   & J_{\mathrm{M}_{(k)}L}:=\\
   &\ ig\!\sum_{n=1}^{d-1}\sqrt{n(d-n)}\Big(\ketbra{n_{\mathrm{M}_{(k)}},n_\mathrm{L}}{n-1_{\mathrm{M}_{(k)}}, n-1_\mathrm{L}}-\mathrm{H.c.}\Big),
   \nonumber
\end{align}
\hl{and 
the} states $\ket{n_{\mathrm{M}_{(k)}}\!}$ are defined as
\begin{align} \label{eq:nNotation}
    \ket{n_{\mathrm{M}_{(k)}}\!} := \bigotimes_{j=1}^{n}|1_\mathrm{C}0_\mathrm{H}\rangle_{\mathrm{M}_k^j}\bigotimes_{l=n+1}^{d-1}|0_\mathrm{C}1_\mathrm{H}\rangle_{\mathrm{M}_k^l}\,.
\end{align}
That is, the state $\ket{n_{\mathrm{M}_{(k)}}}$ can be considered to be the $n^\mathrm{th}$ excited state of the $k^\mathrm{th}$ vertical group M$_{(k)}$ in the sense that the first $n$ machines $\mathrm{M}_{k}^{j}$ for $j=1,\ldots,n$ are in the `used' state $\ket{1_\mathrm{C}0_\mathrm{H}}_{\mathrm{M}_{k}^{j}}$, whereas the remaining $d-n+1$ machines $\mathrm{M}_{k}^{l}$, with $l=n+1,\ldots,d-1$, are in the `unused' state $\ket{0_\mathrm{C}1_\mathrm{H}}_{\mathrm{M}_{k}^{l}}$. \hl{The relative normalisation factor $\sqrt{n(d-n)}$ of the different interaction terms in the Hamiltonian is precisely chosen such that the different subspace rotations are in phase to single out a $\sin^{2(d-1)}(g t)$ scaling of $P_{\mathrm{top}(t)}$ as opposed to a mixture of different powers of trigonometric functions. For further details, see Appendix} \ref{appsec:vertical extension details}.

In the following we will briefly discuss the horizontal and vertical extensions separately to outline their physical impact. 


\vspace*{-2mm}
\subsubsection{Horizontal extension}\label{sec:Horizontal extension}
\vspace*{-2mm}

As shown in Appendix~\ref{appsec: The horizontal extension}, for $T_{\mathrm{C}}=0$, the interaction Hamiltonian for $N=2$ in Eq.~(\ref{generalHamiltonian}) then modifies the top-level probability from Eq.~(\ref{eq:ptop 2qb machine}) to
\begin{align}
    P_{\mathrm{top}}(t) &=\Bigl(1-\frac{1}{\mathcal{Z}_{\mathrm{H}}^{M}}\Bigr)\sin^{2}(g t).
    \label{eq:ptop horizontal}
\end{align}
For finite $T_{\mathrm{C}}$, \hl{the weight of this sinusiodal term changes, and there are additional constant and cosine terms, whose relative weight increase} with increasing $T_{\mathrm{C}}$ (see Appendix~\ref{appsec: The horizontal extension}).

From Eq.~(\ref{eq:ptop horizontal}) we see that the maximal value of $P_{\mathrm{top}}(t)$ increases with increasing $M$, and total population inversion can be achieved in the limit $M\rightarrow\infty$. However, in order to achieve ATPC, only increasing the magnitude of $P_{\mathrm{top}}(t)$ is not sufficient since this neglects the temporal concentration. In the next section we therefore introduce the `vertical' extension, which allows us to temporally concentrate $P_{\mathrm{top}}(t)$, leading to sharper peaks.


\subsubsection{Vertical extension}\label{sec:Vertical extension}

For the vertical extension, we generalise the ladder to a non-degenerate system with $d$ evenly spaced energy eigenstates, with the gap between neighbouring states equal to $E_{\mathrm{L}}$. To each of the $d-1$ pairs of neighbouring energy levels of the vertically extended ladder, a $2$-qubit machine can be coupled in the way described in the previous section. In total, the vertically extended clockwork thus consists of a $d$-dimensional ladder and $d-1$ two-qubit machines, as illustrated in Fig.~\ref{horvert}. The resulting top-level probability for $T_{\mathrm{C}}=0$ becomes
\begin{align}
    P_{\mathrm{top}}(t)= \left(\frac{\mathcal{Z}_{\mathrm{H}}-1}{\mathcal{Z}_{\mathrm{H}}}\right)^{(d-1)}\sin^{2(d-1)}(gt).
\end{align}   

We thus see that just vertically extending the machine makes the temporal distribution sharper, but it also decreases the achievable top-level population.

\subsubsection{General extended clockwork}\label{sec:General extended clock work}
Finally, by combining the horizontal and vertical extension, we can combine the advantages of both, \emph{i.e.} simultaneously increase the top-level population and the sharpness of the temporal distribution. Straightforward calculation of the top-level probability for $T_{\mathrm{C}}=0$ (shown in detail in Appendix~\ref{appsec:vertical extension details}) yields
\begin{align}
    P_{\mathrm{top}}(t)= \Bigl\{1-\Bigl[1-\bigl(\tfrac{\mathcal{Z}_{\mathrm{H}}-1}{\mathcal{Z}_{\mathrm{H}}}\bigr)^{d-1}\Bigr]^M\Bigr\}\sin^{2(d-1)}(gt).
    \label{math:ptop}
\end{align}   
\hl{For $T_{\mathrm{C}}$ only slightly greater than zero, Eq.~(}\ref{math:ptop}) smoothly approximates the above top-level probability (see Appendix~\ref{appsec:vertical extension details}). 
 
A direct consequence of the particular form of the top-level probability in Eq.~(\ref{math:ptop}) is that the amplitude and temporal variance (`sharpness') of $P_{\mathrm{top}}(t)$ can be optimised to within any desired error by controlling the number of machines ($M$) per neighbouring pair of energy levels in the horizontal extension  and the dimension of the ladder ($d$) in the vertical extension, respectively.

Since we restricted the clockwork to consist of qubit machines that have up to $M(d-1)$-body interactions, and the Hilbert space of the machine is $4^{M(d-1)}$-dimensional, the most reasonable quantifier of complexity should be related simply to $M$ and $d$. We therefore focus on elucidating the role of $M$ and $d$ separately.  What we can now see is that, in order to decrease the temporal variance (increasing $d$) while increasing the amplitude (increasing $M$), the complexity necessarily has to increase. Thus, for a fixed complexity there exists a trade-off between temporal variance and probability amplitude. 

In the following sections we will include the irreversible decay mechanism to numerically analyse how accuracy and resolution of clocks are influenced by changes in $M$ and $d$. 


\section{Irreversibility and clock ticks}\label{sec:Irreversibility}

Any autonomous quantum clock (or any clock for that matter) inevitably produces entropy in order to tick~\cite{ErkerMitchisonSilvaWoodsBrunnerHuber2017}, as it needs to be subject to an irreversible evolution. While the internal clockwork produces a temporally well-concentrated and repeating distribution, there needs to be an irreversible process that turns this into a measurable signal.
For this to happen there needs to be a system that is driven out of equilibrium in order to relax back to equilibrium while producing a tick. In our case the system that is driven out of equilibrium is the ladder. \hl{The system with respect to which the ladder is driven out of equilibrium is assumed to couple to the ladder such that the top level is unstable and decays, emitting energy into that system.} \hl{As an example,} \hl{one can take this system to be a photon field at the environment temperature $T_{\mathrm{C}}$ that couples to the ladder, such that when the top level} \hl{population} \hl{decays to the ground state it emits a photon of energy $E_\gamma=(d-1)E_\mathrm{L}$.} \hl{However, note that the assumption that only this particular ladder transition couples to the field is an idealization for}  
\hl{
the purpose of deriving fundamental bounds (as discussed in Sec.~}\ref{sec:framework})\hl{.} 
Since any irreversible process can be viewed as a reversible process on a larger Hilbert space, the presence of such a decay channel \hl{must in principle} also allow for the reverse process of exciting the ladder while absorbing \hl{energy, e.g., in the form of a photon in the example above.} 
However the probability for this to happen can be made arbitrarily small by demanding that the background temperature of the field satisfies $\tfrac{E_\gamma}{k_B T_{\mathrm{C}}}\gg 1$.
 
The number of possible decay processes is vast. However, since our aim is to capture all resources that are necessary to operate a clock, allowing for decay processes that require memory would miss the purpose, since the required resources are not clearly defined for them. We therefore require the photon field to be memoryless, \emph{i.e.} that correlations with the ladder are diluted very quickly and are thus negligible. The resulting dynamics are governed by the law of exponential decay, and thus constitute an ideal case, giving an effective upper bound to the clock performance and allowing us to keep track of the resources that are invested. In particular, the probability density for a tick occurring at time $t$ is given by (see Appendix~\ref{appsec:Tick probability})
\begin{align}
    P_{\mathrm{tick}}(t)=c P_{\mathrm{top}}(t)e^{-c\int P_{\mathrm{top}}(t')dt'},
    \label{eq:ptick}
\end{align}
where $c$ is the coupling strength of the photon field with the top level of the ladder. 

Let us now consider the energetic resources required to run the clock. We first note that, taking $T_\mathrm{C}=0$, as the clockwork state evolves, each branch of its superposition where the ladder's top level is excited corresponds to a transition of $d-1$ machines: $\ket{0_{\mathrm{M}_{(k)}}} \to \ket{d-1_{\mathrm{M}_{(k)}}}$, where the value of $k$ differs between branches, and we recall that the $\lbrace \ket{n_{\mathrm{M}_{(k)}}}\rbrace$ were defined in Eq.~\eqref{eq:nNotation}. Thus, regardless of which branch is realised, if the ladder's top level is excited then the heat flow from the hot bath into the total system is given by $Q_\mathrm{in}=(d-1) E_\mathrm{H}$. This heat flow does the work of driving the ladder from $\ket{0}_\mathrm{L}$ to $\ket{d-1}_\mathrm{L}$, \emph{i.e.} $W=(d-1) E_\mathrm{L}$. After the clock has ticked and the cold qubits of the machines re-thermalise, $Q_\mathrm{out}=(d-1) E_\mathrm{C}$ of heat will be dissipated into the cold bath. Since $E_\mathrm{H}=E_\mathrm{L}+E_\mathrm{C}$, we thus have the usual relation for a thermal machine, \emph{i.e.} $Q_\mathrm{in}=W+Q_\mathrm{out}$, and the thermal efficiency of the process is $\eta_\mathrm{th}:=\frac{W}{Q_\mathrm{in}}=\frac{E_\mathrm{L}}{E_\mathrm{L}+E_\mathrm{C}}$. From this, one can see that as $E_\mathrm{C}/E_\mathrm{L}$ decreases, we approach the Carnot efficiency bound $\eta_\mathrm{th}\leq 1$.

Curiously, for $T_\mathrm{C}=0$ and $M\rightarrow\infty$, the top-level population is just $P_{\mathrm{top}}(t)= \sin^{2(d-1)}(gt)$. If interpreted as a heat engine whose purpose is to charge a battery (the ladder), then one can indeed reach an efficiency of $\eta_\mathrm{th}\approx1$ and still charge the battery in finite time $\tau=\frac{\pi}{2g}$. Even the task of ATPC can be achieved to arbitrary precision at perfect efficiency. One can interpret this as sufficient clockwork complexity permitting perfect efficiency at finite power.

In any case, the efficacy of ATPC and the resulting clock dynamics are essentially determined by the ladder dimension $d$ and the number of driving machines $M$, which together correspond to a simple notion of \emph{clockwork complexity}. In order to investigate how these affect the quality of the clock, we quantify this quality using two notions introduced in~\cite{ErkerMitchisonSilvaWoodsBrunnerHuber2017}. These are the accuracy, which is the average number of ticks until the next tick is off by the average time between two ticks, \emph{i.e.}
\begin{align}
     N=\left(\frac{\overline{t}}{\Delta t}\right)^2\text{, }
     \label{eq:accuracy}
\end{align}
and the resolution, which is the inverse average time between two ticks, \emph{i.e.}
\begin{align}
	R=\frac{1}{\overline{t}}.
	\label{eq:resolution}
\end{align}
Here, $\overline{t}=\int_0^\infty tP_{\mathrm{tick}}(t)dt$ and $(\Delta t)^2=\overline{t^2}-(\overline{t})^2$ with $\overline{t^2}=\int_0^\infty t^2P_{\mathrm{tick}}(t)dt$.

\hl{Let us remark here that our choice of quantifier for the resolution $R$ is not the only option. For instance, another candidate to quantify the resolution would be $\overline{1/t}=\int_0^\infty \tfrac{1}{t}P_{\mathrm{tick}}(t)dt$. However, for this choice, small times would contribute much more strongly to the average than for the inverse of the average times. For example, already a single outlier at a very small value of $t$ would result in a very large value of $\overline{1/t}$, and one would conclude that the resolution was very high, even if most of the events were observed at larger values of $t$. Conversely, choosing $1/\bar{t}$ as a quantifier means that even a few outliers at large values $t$ would result in a low resolution. Therefore, our choice $R=1/\bar{t}$ represents the more conservative of these choices, ensuring that our description results in upper bounds on the resolution.} 

In the following section, we present numerical calculations of the accuracy as a function of the resolution, the clockwork complexity and the energy dissipated per tick.

For comparison, let us take as a baseline an example where there are no qubit machines employed, and the ladder simply begins in equilibrium with the hot bath and emits this energy into the cold bath via the irreversible process, \emph{i.e.} there is no ATPC.
In that case, the top-level probability is constant, \emph{i.e.} $P_{\mathrm{top}}(t)={\exp\left[-\beta_\mathrm{H}(d-1)E_\mathrm{L})\right]/\mathcal{Z}_\mathrm{L}}$, which results in $R=\frac{1}{\overline{t}}=\frac{1}{\Delta t}=c \, {\exp\left[-\beta_\mathrm{H}(d-1)E_\mathrm{L})\right]/\mathcal{Z}_\mathrm{L}}$, and thus the resolution is essentially determined by the decay rate $c$ and the population of the decaying level. The accuracy is simply $N=1$.
This highlights the main purpose of a clockwork: An individual event resulting from pure thermalisation would result in an accuracy of $1$ and come at a work cost of $(d-1)E_{\mathrm{L}}$, whereas the clockwork can increase the accuracy while keeping the work cost of one tick constant.


\section{Numerical results}\label{sec:Numerical results}

\begin{figure}[bp!]
\begin{center}
\includegraphics[width=1\columnwidth]{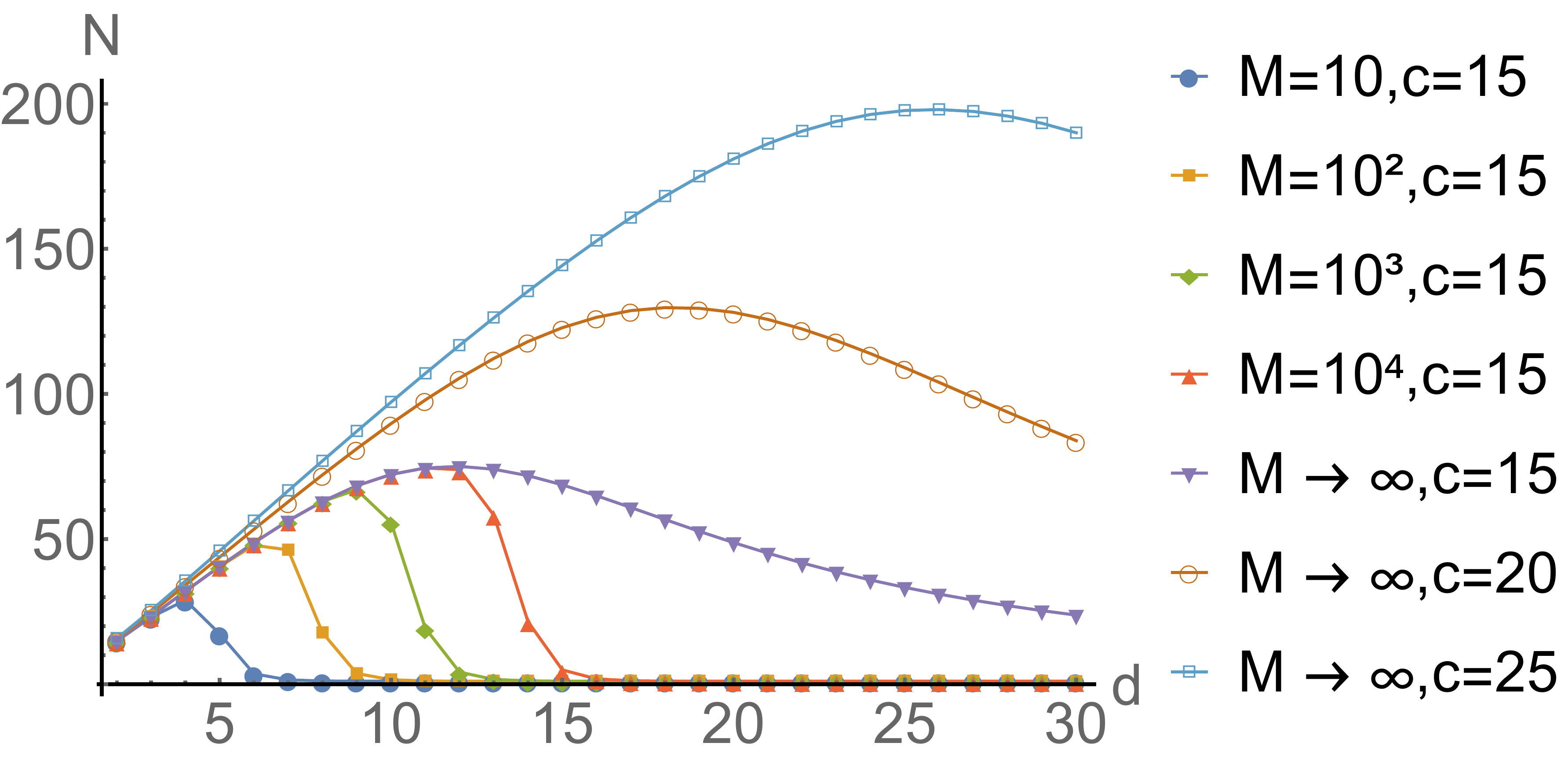}
\caption{The effect of the ladder dimension $d$ on the clock accuracy \hl{$N$} for different $M$. The top three curves show this in the limit $M\rightarrow\infty$ for different $c$, where $[c]=\mathrm{s}^{-1}$.}
\label{fig:AccOverN_nonideal}
\end{center}
\end{figure}

\begin{figure}[tbp]
  \centering
  \includegraphics[width=1\columnwidth]{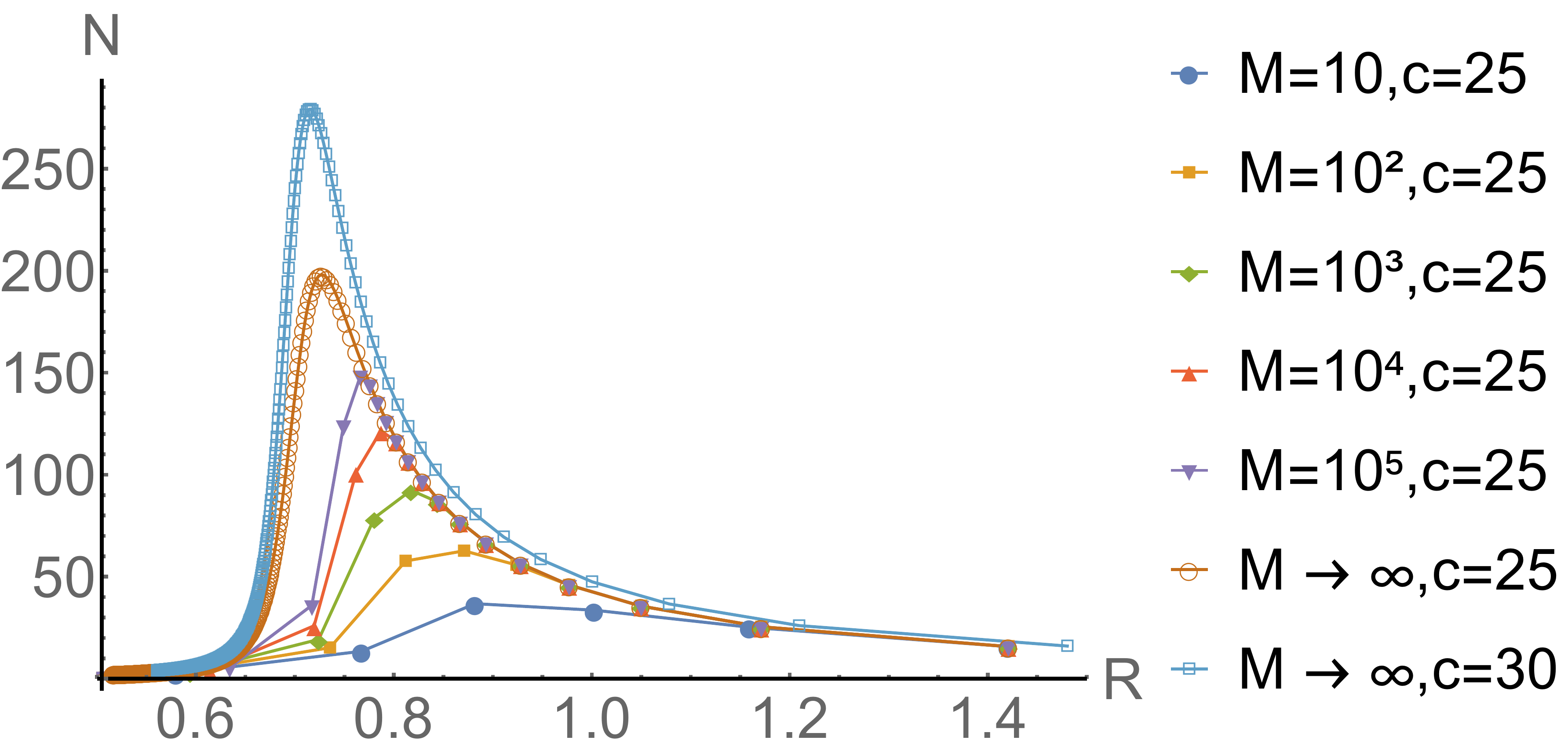}
  \caption{\small{The trade-off between clock accuracy $N$ and resolution $R$ for clockworks of various complexities, where $R$ is increased by decreasing $d$. For finite horizontal extensions $M$, we show this behaviour for a fixed photon field coupling $c=25$, where $[c]=\mathrm{s}^{-1}$. Increasing $M$ allows for higher maximal accuracy, and thus the orange line represents an upper bound of the accuracy for a given resolution for all clocks with $c=25$. For $M\rightarrow\infty$, we see that increasing $c$ increases the potentially achievable combinations accuracy and resolution. We have chosen $g=1~\mathrm{E_C}$ in all cases.}}
  \label{fig:A_R}
\end{figure}

\begin{figure}[tbp]
  \centering
  \includegraphics[width=\columnwidth]{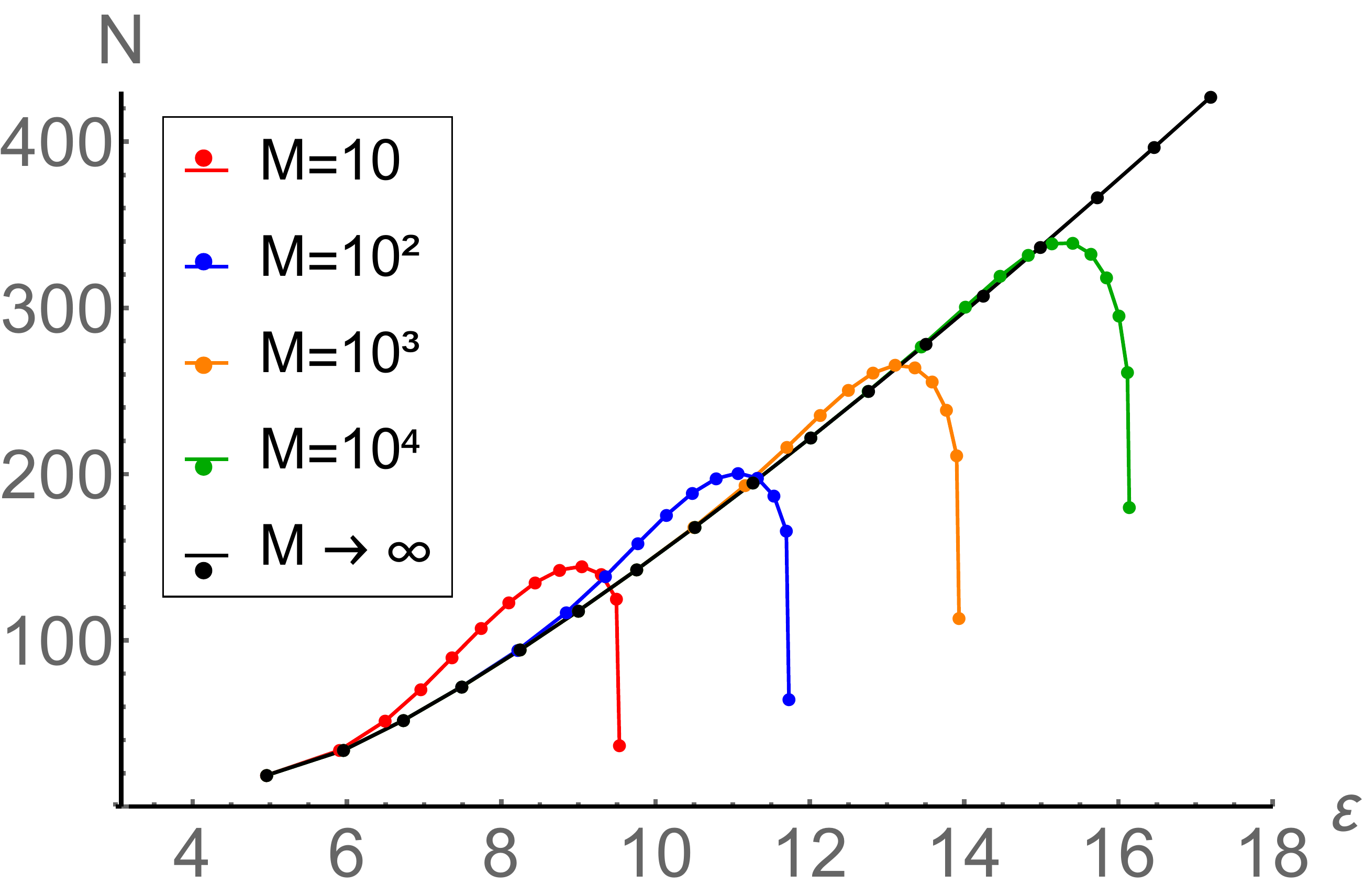}
  \caption{\small{  
  The accuracy \hl{$N$} as a function of the energy dissipation rate $\varepsilon= Q_\mathrm{out} / \bar{t}$ (with $[\varepsilon]=E_\mathrm{C}\, \mathrm{s}^{-1}$) for different numbers of horizontal extensions $M$, \hl{with $c=10^{5}\mathrm{s}^{-1}$.} 
  \protect\hl{As $M$ increases, the maxima of the curves obtained by varying $d$ are shifted towards higher values of $N$ and $\varepsilon$, \emph{i.e.}, up and to the right, reaching finite values (not shown) in the limit $M\rightarrow\infty$ (black curve). The individual peaks obtained for a given coupling strength $c$ and number $M$ of horizontal extensions correspond to the clocks that achieve maximal accuracy under these constraints.} 
  We exclude sub-optimal cases, \emph{i.e.} cases where increasing $d$ reduces resolution and accuracy.
  }}
  \label{fig:acc_Ediss}
\end{figure}

Since we are interested in upper bounds on the clock quality, for the following results we assume the temperature of the hot bath to be infinite, \emph{i.e.} $T_{\mathrm{H}}\rightarrow \infty$, as well as $T_\mathrm{C}=0$. 
\hl{The curves in the following figures} \hl{are generated numerically by varying three free parameters, namely $M$, $d$ and $c$ (varying $g$ has the inverse effect of varying $c$ -- see Appendix}~\ref{appsec:sharpness}\hl{). In particular, each curve corresponds to fixed values of $M$ and $c$ while $d$ varies;} \hl{we display the accuracy $N$ on the vertical axes, and the ladder dimension $d$ (Fig.~}\ref{fig:AccOverN_nonideal}\hl{), the resolution $R$ (Fig.~}\ref{fig:A_R}\hl{), and the energy dissipation rate $\epsilon$ (Fig.~}\ref{fig:acc_Ediss}\hl{), respectively, on the horizontal axes.} 

First we analyse the relation between the `sharpness' of the peak of $P_{\mathrm{top}}(t)$ and the clock accuracy. Recalling the discussion in Sec.~\ref{sec:Vertical extension}, we note that the `sharpness' of $P_{\mathrm{top}}(t)$ increases with increasing $d$, and the latter may therefore be used as a measure of the `sharpness' of $P_{\mathrm{top}}(t)$. In Fig.~\ref{fig:AccOverN_nonideal}, the behaviour of the accuracy as $d$ increases is depicted for different \hl{ladder-bath} coupling strengths $c$ and different values of $M$. For small $d$ we see that the accuracy increases linearly. However, increasing the `sharpness' beyond a certain point leads to a decrease in accuracy \hl{(this behaviour will be discussed in detail in Sec.~}\ref{sec:discussion}). The value of $c$ therefore puts a bound on the maximally achievable accuracy for all potential clocks. The same limiting behaviour is apparent if we fix $c$ and vary $g$ instead (see Appendix~\ref{appsec:sharpness}), for reasons that we discuss below. 

In order to analyse both accuracy and resolution with respect to the clockwork complexity in Fig.~\ref{fig:A_R}, we compare those two quantities for different fixed values of $M$ while varying $d$. Increasing $M$ allows us to reach a higher maximal accuracy, while increasing $d$ (which increases from right to left in Fig.~\ref{fig:A_R}) allows us to trade resolution for accuracy up to the optimal point, after which the accuracy reduces again. We further observe that all clocks with the same $c$ and $g$ lie under a curve defined by the case of $M\rightarrow\infty$. Increasing $c$ allows for clocks of higher quality, \emph{i.e.} that have higher accuracy and resolution. Furthermore, the position of the maximum depends on the value of $g$. Here $g$ was chosen to be equal to $1~\mathrm{E_C}$. Increasing this value shifts the peak to the right, \emph{i.e.} to higher resolutions (see \hl{A}ppendix~\ref{appsec:sharpness}).

Finally, in order to analyse the effect of the energy dissipation rate $\varepsilon= Q_\mathrm{out} / \bar{t}$ on the clock accuracy with respect to \hl{different values of $M$ and $d$, which are simply related to the complexity of the clockwork (see Sec.}~\ref{sec:General extended clock work}), in Fig.~\ref{fig:acc_Ediss} we plot the accuracy over the energy dissipation rate $\varepsilon$ for clockworks of different complexity. In particular, we compare different values of $M$ while varying $d$. What we can see in Fig.~\ref{fig:acc_Ediss} is that for fixed $M$ (at fixed $c$ and $g$), increasing the energy dissipation rate (which is achieved by increasing $d$) increases the accuracy at a certain slope until a maximum is reached. Increasing $d$ further decreases the accuracy. Furthermore, for a given $c$, increasing $M$ leads to a lower slope (approaching the slope of $M\rightarrow\infty$) while allowing for higher maximal accuracy, which suggests that a greater maximal accuracy can be achieved at the cost of a greater energy dissipation rate. \hl{We also note that increasing $c$ increases the maximally achievable accuracy $N$, as can also be seen in Figs.~}\ref{fig:AccOverN_nonideal}  and~\ref{fig:A_R}\hl{. At the same time, increasing $c$ increases the resolution (which can be seen in Fig.}~\ref{fig:acc_Ediss}\hl{). Since $Q_{\mathrm{out}}$ does not depend on $c$, this increase in resolution leads to a higher energy-dissipation rate for all values of $d$ and $M$, thus increasing the slope of the $N$-$\epsilon$ curve (in the linear regime). For reasons of visual clarity this is not depicted in Fig.}~\ref{fig:acc_Ediss}. \hl{One should note, however, that increasing $c$ indefinitely would eventually break the assumptions inherent in our analysis and force us to explore deviations from a Markovian exponential decay towards memory effects in the irreversible dynamics.}


\section{Discussion}\label{sec:discussion}

Our results have two general implications. The first concerns the task of autonomous probability concentration. We show that, in principle, sufficiently increasing the clockwork complexity alone is enough to concentrate the temporal probability arbitrary well. In particular, this task can again be split into two conceptually different sub-tasks: maximizing the achievable population and improving the temporal `sharpness' with which this (maximal) population can be reached. By splitting the clockwork into a target ladder and virtual-qubit machines coupled to the different ladder transitions, we were able to analyse how more complex clockworks can help to achieve the two respective sub-tasks. While we have worked with equally-spaced ladder systems, the same result (qualitatively) also holds for arbitrarily spaced target Hamiltonians, simply by redefining the respective coupling strengths (the $g$'s) of the interaction Hamiltonians. We have equipped our clockwork with a particular tensor product structure, the division into two-qubit machines, for the sake of keeping track of its complexity. Our machine operates optimally within the framework set by this division, but whether more general machines could also achieve the same performance with smaller overall size remains an open question.

In our analysis, we have optimised the internal structure of the clock, \emph{i.e.} the clockwork, to concentrate the probability in a fashion that most closely resembles the temporal distribution of an ideal clockwork. For given $c$, this amplifies the clock quality only up to a limit, which we showcase in Fig.~\ref{fig:AccOverN_nonideal}. This can intuitively be understood by considering the two key timescales of the clock, namely that of the clockwork's dynamics, and that of the irreversible decay. Increasing $d$ while keeping $c$ fixed, one eventually reaches a point where $P_{\mathrm{top}}(t)$ is so well concentrated temporally that the comparative slowness of the decay mechanism reduces the probability that the clock will tick. In other words, it becomes more likely that the decay mechanism will skip that peak. We see the inverse of this behaviour if instead of $c$, we consider curves of fixed $g$ (see Appendix~\ref{appsec:sharpness}), as increasing $g$ speeds up the clockwork, effectively making the limit imposed by $c$ more restrictive.

This brings us to the second \hl{implication of our work. The} irreversible mechanism, in our case characterised by the parameter $c$, puts an absolute upper bound on the achievable combinations of resolution and accuracy, \emph{i.e.} the clock quality,  and thus determines the potential for how well a particular physical process can be used as the basis for a clock. The question of how well this upper bound can be approximated brings us to the role of our two extensions. First of all, the horizontal extension, \emph{i.e.} the coupling of multiple elementary machines to a single transition  between neighbouring ladder levels primarily serves the purpose of increasing the possible population inversion and with it the achievable top-level population. As we see in Eq.~(\ref{eq:ptick}), $c$ always appears multiplied by the prefactor of the sine in Eq.~(\ref{math:ptop}), resulting in an \emph{effective coupling}:
\begin{align} \label{eq:effectivec}
    C_M=c\left\{1-\left[1-\left(\dfrac{\mathcal{Z}_{\mathrm{H}}-1}{\mathcal{Z}_{\mathrm{H}}}\right)^{d-1}\right]^M\right\}\,.
\end{align} 
From this we can see that increasing the horizontal extension $M$ is physically equivalent to increasing the coupling $c$, and this is why they play the same role in Fig.~\ref{fig:AccOverN_nonideal} (though we note that $C_M$ is bounded with respect to $M$ but not with respect to $c$). One cannot make a similar statement to relate $c$ with the vertical extension $d$, since the exponent of the sine in Eq.~(\ref{math:ptop}) will vary as $d$ does. As noted above, $d$ sharpens the temporal distribution, thus increasing the accuracy of the clock as long as the limit set by $c$ and $M$ \hl{[}via Eq.~\eqref{eq:effectivec}\hl{]} is not surpassed, as demonstrated in Fig.~\ref{fig:AccOverN_nonideal}. 

In the regime where the accuracy grows linearly with the `sharpness' (Fig.~\ref{fig:AccOverN_nonideal}), which is determined by $d$ (see Appendix~\ref{appsec:sharpness}), there exists a trade-off relation between accuracy and resolution. To see this, note that the resolution decreases monotonically with $d$ \hl{[see Fig.~}\ref{fig:sharpness}(b)]. For fixed $c$ and $g$, the case of $M\rightarrow\infty$ represents an upper bound on the clockwork quality, \emph{i.e.} on the achievable combinations of accuracy and resolution, which is illustrated in Fig.~\ref{fig:A_R}.

In our computations, we have focused on the limit $T_\mathrm{C}\rightarrow 0$. State-of-the-art atomic clocks operate at optical frequencies~\cite{RevModPhys.87.637}, and at even higher frequencies in novel proposals~\cite{Seiferle_2019}. The vacuum state of any optical-frequency mode of the electromagnetic field has a population of $\approx 1$ at room temperature, and this situation is thus virtually indistinguishable from temperature $0$. For clocks operating at much lower frequencies, such as those based on microwave transitions, cooling the environmental degrees of freedom into which the irreversible mechanism dissipates heat would be necessary in order to approach the fundamental limits we derive here.

It is nonetheless important to stress that we are interested in fundamental limits of timekeeping and the associated complexity and cost\hl{, which one of the reasons why we consider autonomous clocks in contrast with atomic clocks, for example, which require external control.} For the practical purpose of building clocks for everyday use, atomic clocks can require as little as $30$~mW~\cite{lutwak2004chip}, which is many orders of magnitude above the scale defined by the system energy and the timescale of the relevant processes, but still insignificant for global energy use.  The majority of that cost, however, comes from the fact that at some point that single event needs to be amplified and registered by a measurement apparatus, whose inherent irreversible nature is also thermodynamic and comes with its own costs and limitations~\cite{GuryanovaFriisHuber2020}. Conversely, the inevitable imperfections of clocks and the associated costs also limit the achievable quality of measurements and, consequently, of all estimation procedures, e.g.\hl{,} of work itself~\cite{DebarbaEtAl2019}.

\hl{As far as current and future prospects are concerned the limits derived in this paper have more fundamental} relevance for the autonomous control of quantum systems by a quantum clock~\cite{Malabarba_2015,ball2016role}. Here, a small quantum system is envisioned to be controlled by an autonomous quantum clock. This is important for any type of unitary process requiring precise timing, from small machines operating in cycles~\cite{Feldmann} to general repeating unitary processes such as circuit-based cooling models~\cite{RaeisiPRL2019,Alhambra_2019,Clivaz_2019L,Clivaz_2019E,Rodriguez-Briones2016}.

Coming back to the actual energy cost, there are a number of interesting observations that follow from our clock model. First of all, \hl{the horizontal extension always comes at a finite energy cost and dissipation for each tick for any clock, while the vertical extensions linearly increase the energy cost and dissipation}. The vertical extensions on the other hand linearly increases the energy cost and dissipation and, as long as the limit imposed by the $c$ and $M$ is not exceeded, also increases the accuracy. \hl{This limit can be qualitatively understood as a point after which further concentrating probability is actually detrimental to clock performance. We can observe that the increase in accuracy follows a linear behaviour in $d$, before switching to a sublinear behaviour close to the peak, after which it actually decreases accuracy again. Thus, in addition to recovering the observation that }$N\propto \Delta S$, \emph{i.e.} that a clock's accuracy is essentially determined by the entropy it dissipates (which seems to be a prevalent feature in all classical and quantum clocks \cite{PearsonEtal2020}), \hl{we have  pinpointed which combination of the resources $M$, $d$ and $c$ allow us to maintain this linear regime. The proportionality factor itself can be identified numerically from the plots in Fig.}~\ref{fig:acc_Ediss}.

Finally, let us put our clockwork model in context with recent literature on quantum clocks. In Refs.~\cite{s2015quantum,s2018performance,WoodsSilvaPuetzStuparRenner2018,yang2020ultimate}, the relationship between achievable clock accuracy as a function of 'clock dimension' is studied by means of repeated applications of maps from a clock system to a register. These works provide fundamental bounds for clock accuracy for fixed system dimension $d$, showing that the accuracy of classical (incoherent) clocks can at best scale linearly in $d$, whereas a quantum clock's accuracy (with states featuring coherence) may scale as $d^2$. The clock system considered in these works is exactly what we here refer to as the ladder system. Meanwhile, the map that Refs.~\cite{s2015quantum,s2018performance,WoodsSilvaPuetzStuparRenner2018,yang2020ultimate,woods2020autonomous} refer to as being responsible for creating a tick event in the register subsumes the interactions between the ladder, the qubit machines, and the heat baths, and also includes the irreversible mechanism and the subsequent read-out. In other words, our work provides a concrete physical realisation of the maps that effect the transfer of ticks to the register. In Fig.~\ref{fig:AccOverN_nonideal}, we see that, in the regime of the clockwork not exceeding the clock potential dictated by the irreversible mechanism, the accuracy scales linearly with the ladder dimension $d$, which is already the optimal achievable scaling~\cite{yang2020ultimate}.


\section{Conclusion}\label{sec:conc}

In this article, we have put forward a framework for studying fundamental limits of timekeeping. The conceptual split of any such task into a \emph{clockwork}, which creates a temporally concentrated probability distribution, and a mechanism for irreversibility, allowed us to derive an analytic formula for the achievable temporal probability concentration of the clockwork. The irreversible mechanism provides a context for the operation of the clock by allowing the passage of time to be tracked in the first place. Meanwhile, the chosen irreversible mechanism sets the reference timescale that ultimately constrains the potential of any clockwork that harnesses this mechanism to form a clock. 
But it is the clockwork that needs to be appropriately tuned to achieve maximal performance given these constraints. By composing the clockwork of the smallest possible thermal machines, we were further able to conceptually split the task of autonomous probability concentration into two sub-tasks. First, by having more machines interact with a single transition, we can increase the maximum top-level probability and with it, the effective coupling to the irreversible mechanism. Second, by concatenating multiple transitions of this kind, we are able to sharpen the temporal distribution. This reveals the intricate ways in which the complexity of the clockwork determines its performance.

\hl{In the future, it will be interesting to study more exotic irreversible mechanisms beyond exponential decay and whether they could be harnessed to further improve the clock quality. Moreover, one might study increasing clockwork complexity in a manner other than the addition of qubits. Furthermore, beyond simply asking if optimal ATPC is achievable in some limit, it would be of interest to ask which temporal probability profile (\textit{i.e.} $P_{\mathrm{tick}}(t)$) optimizes clock performance under some constraints, and the extent to which such a profile is achievable with a quantum machine. Another interesting avenue would be to make a stronger connection to experimental implementations of clocks and how the notion of irreversibility versus temporal probability concentration can be more formally made on larger scales. One experiment in that direction is performed in Ref.}~\cite{PearsonEtal2020}\hl{, using a nano-mechanical membrane. Photoisomerisation could also provide a potential platform for implementing such a clock at molecular scales} \cite{alex3}\hl{. While there are many open paths to explore and questions to answer, our results consolidate the fact that perfect clocks are practically impossible when derived from first principles and that significant thermodynamic resources have to be invested to reach the potential of any physical system to act as a clock.}\\ 

\emph{Acknowledgements}.
We acknowledge support from the Austrian Science Fund (FWF) through the START project Y879-N2, the Zukunftskolleg ZK03 and the project P 31339-N27. We further acknowledge funding from the ESQ Discovery grant ``\emph{Emergence of physical laws: From mathematical foundations to applications in many body physics}''.


\bibliographystyle{apsrev4-1fixed_with_article_titles_full_names_new}
\bibliography{bibfile}


\newpage
\hypertarget{sec:appendix}
\onecolumngrid
\appendix

\renewcommand{\thesubsection}{\thesection.\arabic{subsection}}
\renewcommand{\thesubsubsection}{\thesubsection.\arabic{subsubsection}}

\section*{Appendices}

In these appendices, we provide detailed derivations and background information for the results presented in the main text. In Appendix~\ref{appesec:top-level probability of a two-qubit clockwork} we first present a derivation of the top-level probability for the two-qubit machine of Sec.~\ref{Two-Qubit Machine}. In Appendix~\ref{appsec: The horizontal extension}, we then present the derivation of the top-level probability for the horizontal extension of the clockwork for arbitrary temperatures $T_{\mathrm{H}}$ and $T_{\mathrm{C}}$. In Appendix~\ref{appsec:vertical extension details}, we then again focus on the case $T_{\mathrm{C}}=0$, for which we derive the top-level probability in the full horizontal and vertical extension. Appendix~\ref{appsec:Tick probability} contains the derivation of the tick probability. Appendix~\ref{appsec:numerical} presents the details for the numerical computation of accuracy and resolution. Appendix~\ref{appsec:sharpness} discusses the behaviour of clocks with changing ladder dimension $d$ as well as changing coupling constant $g$. 

\section{Top-level probability of a two-qubit clockwork}
\label{appesec:top-level probability of a two-qubit clockwork}

Here, we present a derivation of the top-level probability $P_{\mathrm{top}}(t)$ from Eq.~(\ref{eq:ptop 2qb machine}). That is, we consider the minimal clockwork discussed in Sec.~\ref{Two-Qubit Machine}, which consists of a hot qubit (coupling to the hot bath at temperature $T_\mathrm{H}$, as well as a cold qubit and a ladder (both coupling to a cold bath at temperature $T_{\mathrm{C}}$). The derivation presented here is a special case ($M=1$ and $T_{\mathrm{C}}=0$) of the more general derivation of the top-level probability within the horizontal extension that we will present in Appendix~\ref{appsec: The horizontal extension} (where $M>1$ and both $T_\mathrm{H}$ and $T_{\mathrm{C}}$ can take on arbitrary values). Nevertheless, we will first go through the much simpler derivation for $M=1$ and $T_{\mathrm{C}}=0$ here, which will serve as a guiding example for the much more involved general calculation that is to follow.

Assuming that the systems have thermalised with their respective baths, the initial state of the clockwork is given by
\begin{align}
    \rho_{0} &=\, \ket{0}\!\!\bra{0}_\mathrm{C}\otimes\tau_\mathrm{H}\otimes\ket{0}\!\!\bra{0}_\mathrm{L}\,=\,
    \ket{0}\!\!\bra{0}_{\mathrm{C}}\otimes
    \tfrac{1}{\mathcal{Z}_\mathrm{H}}\bigl(\ket{0}\!\!\bra{0}_{\mathrm{H}}+
    e^{-\beta_{\mathrm{H}}E_{\mathrm{H}}}\,\ket{1}\!\!\bra{1}_{\mathrm{H}}
    \bigr)
    \otimes\ket{0}\!\!\bra{0}_{\mathrm{L}},
\end{align}
where $\mathcal{Z}_{\mathrm{H}} = 1+ e^{-\beta_{\mathrm{H}}E_{\mathrm{H}}}$ is the partition function of the hot qubit. Since the interaction term $H_{\mathrm{int}}$ in the total Hamiltonian $H=H_{0}+H_{\mathrm{int}}$ is chosen such that the free energy (that is, w.r.t. the free Hamiltonian $H_{0}$)  is conserved, $\left[H_{0},H_{\mathrm{int}}\right]=0$, and because the initial state $\rho_{0}$ is diagonal in the eigenbasis of $H_{0}$, we can write the top-level probability from Eq.~(\ref{eq:def top level prob}) as
\begin{align}
    P_{\mathrm{top}}(t) &=
    \tr\bigl(\ket{1}\!\!\bra{1}_{\mathrm{L}} e^{-iHt}\rho_{0}\,e^{iHt}\bigr)
    \,=\,
    \tr\bigl(\ket{1}\!\!\bra{1}_{\mathrm{L}} e^{-iH_{\mathrm{int}}t}\rho_{0}\,e^{iH_{\mathrm{int}}t}\bigr)
    \,=\,
    \tr\bigl(T_{0} \ket{0}\!\!\bra{0}_\mathrm{C}\otimes\tau_\mathrm{H} T_{0}^{\dagger}\bigr),
    \label{eq:def top level prob appendix}
\end{align}
where $T_{0}$ is a matrix encoding the transition amplitude between ground state and excited state of the ladder, given by
\begin{align}
    T_{0} & =\,_{\mathrm{L}}\!\!\bra{1}e^{-iH_{\mathrm{int}}t}\ket{0}_{\mathrm{L}}
    \,=\,
    \sum\limits_{k=0}^{\infty}\frac{(-it)^{k}}{k!}\,
    \,_{\mathrm{L}}\!\!\bra{1}H_{\mathrm{int}}^{k}\ket{0}_{\mathrm{L}}.
    \label{eq:def top level amplitude appendix}
\end{align}
Now, because $H_{\mathrm{int}}^{2}$ is proportional to the identity on its support, that is, 
\begin{align}
    H_{\mathrm{int}}^{2} & =\,g^{2}\,\bigl(
    \ket{1}\!\!\bra{0}_{\mathrm{C}}\otimes\ket{0}\!\!\bra{1}_{\mathrm{H}}\otimes\ket{1}\!\!\bra{0}_{\mathrm{L}}\,+\,\mathrm{H.c.}
    \bigr)^{2}
    \,=\,
    g^{2}\,\bigl(
    \ket{1}\!\!\bra{1}_{\mathrm{C}}\otimes\ket{0}\!\!\bra{0}_{\mathrm{H}}\otimes\ket{1}\!\!\bra{1}_{\mathrm{L}}\,+\,\ket{0}\!\!\bra{0}_{\mathrm{C}}\otimes\ket{1}\!\!\bra{1}_{\mathrm{H}}\otimes\ket{0}\!\!\bra{0}_{\mathrm{L}}
    \bigr),
    \label{eq:square is identity on support}
\end{align}
even powers of $H_{\mathrm{int}}$ do not contribute to $T_{0}$. However, for odd powers we have $H_{\mathrm{int}}^{2k+1}=g^{2k}H_{\mathrm{int}}$, such that we find $\,_{\mathrm{L}}\!\!\bra{1}H_{\mathrm{int}}^{2k+1}\ket{0}_{\mathrm{L}}=g^{2k+1}\ket{1}\!\!\bra{0}_{\mathrm{C}}\otimes\ket{0}\!\!\bra{1}_{\mathrm{H}}$. With this, we can evaluate the transition matrix $T_{0}$, \emph{i.e.} 
\begin{align}
    T_{0} & =\,
    \sum\limits_{k=0}^{\infty}\frac{(-it)^{2k+1}}{(2k+1)!}\,
    \,_{\mathrm{L}}\!\!\bra{1}H_{\mathrm{int}}^{2k+1}\ket{0}_{\mathrm{L}}
    \,=\,
    \sum\limits_{k=0}^{\infty}\frac{(-igt)^{2k+1}}{(2k+1)!}\,
    \ket{1}\!\!\bra{0}_{\mathrm{C}}\otimes\ket{0}\!\!\bra{1}_{\mathrm{H}}
    \,=\,\sin(gt)\,\ket{1}\!\!\bra{0}_{\mathrm{C}}\otimes\ket{0}\!\!\bra{1}_{\mathrm{H}}.
    \label{eq:top level amplitude 2qb machine}
\end{align}
Inserting the result into Eq.~(\ref{eq:def top level prob appendix}), we finally arrive at the top-level probability
\begin{align}
    P_{\mathrm{top}}(t) &=\,
    \tr\bigl(T_{0} \ket{0}\!\!\bra{0}_\mathrm{C}\otimes\tau_\mathrm{H} T_{0}^{\dagger}\bigr)
    \,=\,\sin^{2}(gt)\,\,_{\mathrm{L}}\!\!\bra{1}\tau_{\mathrm{H}}\ket{0}_{\mathrm{L}}
    \,=\,\sin^{2}(gt)\,\frac{e^{-\beta_{\mathrm{H}}E_{\mathrm{H}}}}{1+e^{-\beta_{\mathrm{H}}E_{\mathrm{H}}}}
    \,=\,\sin^{2}(gt)\,\Bigl(1-\frac{1}{\mathcal{Z}_{\mathrm{H}}}\Bigr).
    \label{eq:top level prob 2qb machine appendix}
\end{align}


\section{The horizontal extension}
\label{appsec: The horizontal extension}

In this appendix, we present more technical details of the horizontal extension of the autonomous clockwork from a single ($M=1$) to multiple ($M>1$) two-qubit machines interacting with the same two-level ($d=2$) transition of the ladder system. In particular we will derive the top-level probability for the horizontal extension for arbitrary temperatures $T_\mathrm{C}$ and $T_\mathrm{H}$.

Following a similar approach as in Eq.~(\ref{eq:def top level prob appendix}), we define transition operators
\begin{align}
    &T_{n}:=\;_\mathrm{L}\!\!\bra{1}e^{-i H_{\mathrm{int}} t}\ket{n}_\mathrm{L} =\;_\mathrm{L}\!\!\bra{1}\sum_{j=0}^{\infty}\dfrac{(-i t)^j}{j!} H_{\mathrm{int}}^j\ket{n}_\mathrm{L} 
    =\;_\mathrm{L}\!\!\bra{1}\sum_{j=0}^{\infty}\dfrac{(-i t)^j}{j!}\sum_{k=1}^{M} H_{k}^j\ket{n}_\mathrm{L},
\end{align}
for $n=0,1$, where the last equality follows from the fact that the interaction terms $H_{k}$ given by the terms in Eq.~(\ref{generalHamiltonian}) have mutually disjoint support, \textit{i.e.}  $H_{k} H_{k'}=0$ for $k \neq k'$. Before we calculate these transition operators, we note that $H_{k}$ satisfies the cyclic property
\begin{align}
    H_{k}^{2q} &= g^{2q} \bigotimes_{i=1}^{k-1}\mathds{1}_{\mathrm{M}_i}\otimes\left( \ket{0_\mathrm{C}1_\mathrm{H}}\!\!\bra{0_\mathrm{C}1_\mathrm{H}}_{\mathrm{M}_{k}}\otimes \ket{0}\!\!\bra{0}_{\mathrm{L}} + \ket{1_\mathrm{C}0_\mathrm{H}}\!\!\bra{1_\mathrm{C}0_\mathrm{H}}_{\mathrm{M}_{k}}\otimes\ket{1}\!\!\bra{1}_{\mathrm{L}} \right)\otimes\bigotimes_{j=k+1}^{M}{\Pi_{\mathrm{M}_j}}  &\text{for }q\in\mathds{N}_{>0}, \\
    H_{k}^{2q+1} &= g^{2q+1} \bigotimes_{i=1}^{k-1}\mathds{1}_{\mathrm{M}_i}\otimes \left( \sigma_{\mathrm{M}_{k}}^{-}\otimes\sigma_{\mathrm{L}}^{+} + \sigma_{\mathrm{M}_{k}}^{+}\otimes\sigma_{\mathrm{L}}^{-} \right)\otimes\bigotimes_{j=k+1}^{M}\Pi_{\mathrm{M}_j}\,=\, g^{2q+1} H_{k} &\text{for }q\in\mathds{N}\;\;\;\;, 
\end{align}
where \hl{$\mathrm{M}_i=\mathrm{M}^1_i=\mathrm{C}^1_i\otimes\mathrm{H}^1_i$ denotes the Hilbert space of the $i^\mathrm{th}$ horizontal extension,} $\sigma^{+}_\mathrm{L}:=\ket{1_\mathrm{L}}\!\!\bra{0_\mathrm{L}}=\left(\sigma^{-}_\mathrm{L}\right)^\dag$, $\sigma_{\mathrm{M}_{k}}^{+}:=\ket{0_\mathrm{C}1_\mathrm{H}}\!\!\bra{1_\mathrm{C}0_\mathrm{H}}_{\mathrm{M}_k}=\left(\sigma_{\mathrm{M}_{k}}^{+}\right)^\dag$, and $\ket{l_\mathrm{C}m_\mathrm{H}}\!\!\bra{p_\mathrm{C}q_\mathrm{H}}_{\mathrm{M}_k}:=\ket{l}\!\!\bra{p}_{\mathrm{C}_k}\otimes\ket{m}\!\!\bra{q}_{\mathrm{H}_k}$, with $l,m,p,q=0,1$.

Now, for the transition operator $T_{0}$, we note that only odd powers of $H_{k}$ can map $\ket{0}_{\mathrm{L}}$ to $\ket{1}_{\mathrm{L}}$, and therefore
\begin{align}
    T_{0}=\sum_{q=0}^{\infty}\sum_{k=1}^{M}\dfrac{(-i t)^{2q+1}}{(2q+1)!}\;_\mathrm{L}\!\!\bra{1}H_{k}^{2q+1} \ket{0}_\mathrm{L}=\sin(g t)\left(\sum_{k=1}^{M} \bigotimes_{j=1}^{k-1}\mathds{1}_{\mathrm{M}_j}\otimes \sigma_{\mathrm{M}_k}^{-}\otimes \bigotimes_{l=k+1}^{M}\Pi_{\mathrm{M}_l}\right).
    \label{eq:Atop horizontal}
\end{align}
We can calculate $T_{1}$ similarly, noting that only even powers of $H_{k}$ contain the factor $\ket{1}\!\!\bra{1}_\mathrm{L}$.
\begin{align}
     &T_{1}=\sum_{q=0}^{\infty}\sum_{k=1}^{M}\dfrac{(-i t)^{2q}}{(2q)!}\;_\mathrm{L}\!\!\bra{1}H_{k}^{2q} \ket{1}_\mathrm{L}  =\cos(g t)\left(\sum_{k=1}^{M}\bigotimes_{j=1}^{k-1}\mathds{1}_{\mathrm{M}_j}\otimes \ket{1_\mathrm{C}0_\mathrm{H}}\!\!\bra{1_\mathrm{C}0_\mathrm{H}}_{\mathrm{M}_k}\otimes \bigotimes_{l=k+1}^{M}\Pi_{\mathrm{M}_l}\right) + \tilde{\Pi} \label{eq:T1horizontal}
\end{align}
where $\tilde{\Pi}$ is a projection defined by
\begin{align}
    \tilde{\Pi}:=\mathds{1}_{\mathcal{H}\backslash\mathrm{L}}-\sum_{k=1}^{M}\bigotimes_{j=1}^{k-1}\mathds{1}_{\mathrm{M}_j}\otimes \ket{1_\mathrm{C}0_\mathrm{H}}\!\!\bra{1_\mathrm{C}0_\mathrm{H}}_{\mathrm{M}_k}\otimes \bigotimes_{l=k+1}^{M}\Pi_{\mathrm{M}_l}=\bigotimes_{k=1}^{M}\Pi_{\mathrm{M}_k}+\sum_{k=1}^M\bigotimes_{j=1}^{k-1}\mathds{1}_{\mathrm{M}_{j}}\otimes \ket{0_\mathrm{C}1_\mathrm{H}}\!\!\bra{0_\mathrm{C}1_\mathrm{H}}_{\mathrm{M}_k}\bigotimes_{j'=k+1}^M \Pi_{\mathrm{M}_{j'}}.
\end{align}

In order to evaluate the top-level probability, let us briefly inspect the initial state $\rho_0$ in this situation, which is given by
\begin{align}
   \rho_{0}& =\,\bigotimes_{i=1}^{M}\tau_{\mathrm{C}_i}\otimes\tau_{\mathrm{H}_i}\otimes\tau_{\mathrm{L}}\nonumber\\
   &=\underbrace{\bigotimes_{i=1}^{M}\left(\dfrac{1}{\mathcal{Z}_{\mathrm{C}}}\ket{0}\!\!\bra{0}_{\mathrm{C}_i}+\dfrac{\mathcal{Z}_{\mathrm{C}}-1}{\mathcal{Z}_{\mathrm{C}}}\ket{1}\!\!\bra{1}_{\mathrm{C}_i} \right)\otimes\left(\dfrac{1}{\mathcal{Z}_{\mathrm{H}}}\ket{0}\!\!\bra{0}_{\mathrm{H}_i}+\dfrac{\mathcal{Z}_{\mathrm{H}}-1}{\mathcal{Z}_{\mathrm{H}}}\ket{1}\!\!\bra{1}_{\mathrm{H}_i} \right)}_{\tr_{\mathrm{L}}(\rho_0)}\otimes\left(\dfrac{1}{\mathcal{Z}_{\mathrm{L}}}\ket{0}\!\!\bra{0}_{\mathrm{L}}+\dfrac{\mathcal{Z}_{\mathrm{L}}-1}{\mathcal{Z}_{\mathrm{L}}}\ket{1}\!\!\bra{1}_{\mathrm{L}} \right) .
\end{align}
We split the initial state into two parts: one where the ladder is initially exited and one where it is not, \emph{i.e.}
\begin{align}
    \rho_{0}=\dfrac{1}{\mathcal{Z}_{\mathrm{L}}}\tr_{\mathrm{L}}(\rho_0)\otimes\ket{0}\!\!\bra{0}_{\mathrm{L}}+\dfrac{\mathcal{Z}_{\mathrm{L}}-1}{\mathcal{Z}_{\mathrm{L}}}\tr_{\mathrm{L}}(\rho_0)\otimes\ket{1}\!\!\bra{1}_{\mathrm{L}} .
    \label{hor ext split initial state}
\end{align}
The top-level probability can then be seen to split into two be separate contributions, corresponding to the two terms in Eq.~(\ref{hor ext split initial state}), that is, 
\begin{align}
   P_{\mathrm{top}}(t)=\dfrac{1}{\mathcal{Z}_{\mathrm{L}}}\tr\bigl(T_{0} \tr_{\mathrm{L}}(\rho_0) T_{0}^{\dagger}  \bigr)\, +\,
    \dfrac{\mathcal{Z}_{\mathrm{L}}-1}{\mathcal{Z}_{\mathrm{L}}}\tr\bigl(T_{1} \tr_{\mathrm{L}}(\rho_0) T_{1}^{\dagger}  \bigr), 
\end{align}
We will first consider the part of $\rho_0$ where the ladder is initially in the ground state. Considering  $\bigotimes_{j=1}^{k-1}\mathds{1}_{\mathrm{M}_j}\otimes \sigma_{\mathrm{M}_k}^{-}\otimes \bigotimes_{l=k+1}^{M}\Pi_{\mathrm{M}_l}$ in Eq.~(\ref{eq:Atop horizontal}) we see that for each $k$ there are $k-1$ machines that are acted upon only by identities, meaning the partial trace over each of these machines simply contributes a factor $1$. There are $M-k$ machines that are acted upon by an operator $\Pi_{\mathrm{M}_j}$, each leading to a factor
\begin{align}
 &   \tr\left[\dfrac{1}{\mathcal{Z}_{\mathrm{H}}\mathcal{Z}_{\mathrm{C}}}\Pi_{\mathrm{M}_j}\left(\ket{0}\!\!\bra{0}_{\mathrm{C}_j}+(\mathcal{Z}_{\mathrm{C}}-1)\ket{1}\!\!\bra{1}_{\mathrm{C}_j} \right)\otimes\left(\ket{0}\!\!\bra{0}_{\mathrm{H}_j}+(\mathcal{Z}_{\mathrm{H}}-1)\ket{1}\!\!\bra{1}_{\mathrm{H}_j}\right)\Pi_{\mathrm{M}_j}^{\dagger}\right]\nonumber\\
    &\ =\tr\left[\dfrac{1}{\mathcal{Z}_{\mathrm{H}}\mathcal{Z}_{\mathrm{C}}}\left(\ket{0_\mathrm{C}0_\mathrm{H}}\!\!\bra{0_\mathrm{C}0_\mathrm{H}}_{\mathrm{M}_j}+(\mathcal{Z}_{\mathrm{H}}-1)(\mathcal{Z}_{\mathrm{C}}-1)\ket{1_\mathrm{C}1_\mathrm{H}}\!\!\bra{1_\mathrm{C}1_\mathrm{H}}_{\mathrm{M}_j}\right) \right]=\dfrac{1+(\mathcal{Z}_{\mathrm{C}}-1)(\mathcal{Z}_{\mathrm{H}}-1)}{\mathcal{Z}_{\mathrm{H}}\mathcal{Z}_{\mathrm{C}}} ,
\end{align}
in $\tr\bigl(T_{0} \tr_{\mathrm{L}}(\rho_0) T_{0}^{\dagger}  \bigr)$. In addition, there is always exactly one machine which is acted upon by $\sigma_{\mathrm{M}_k}^{-}$, contributing a factor of 
\begin{align}
    &\tr\left[ \dfrac{1}{\mathcal{Z}_{\mathrm{H}}\mathcal{Z}_{\mathrm{C}}}\sigma_{\mathrm{M}_k}^{-}\left(\ket{0}\!\!\bra{0}_{\mathrm{C}_k}+(\mathcal{Z}_{\mathrm{C}}-1)\ket{1}\!\!\bra{1}_{\mathrm{C}_k} \right)\otimes\left(\ket{0}\!\!\bra{0}_{\mathrm{H}_k}+(\mathcal{Z}_{\mathrm{H}}-1)\ket{1}\!\!\bra{1}_{\mathrm{H}_k}\right)\sigma_{\mathrm{M}_k}^{+} \right]\nonumber\\
    &\ = \tr\left[ \dfrac{1}{\mathcal{Z}_{\mathrm{H}}\mathcal{Z}_{\mathrm{C}}} (\mathcal{Z}_{\mathrm{H}}-1)\ket{1_\mathrm{C}0_\mathrm{H}}\!\!\bra{1_\mathrm{C}0_\mathrm{H}} \right]=\dfrac{\mathcal{Z}_{\mathrm{H}}-1}{\mathcal{Z}_{\mathrm{H}}\mathcal{Z}_{\mathrm{C}}} \label{eq:OneMachineFactor}
\end{align}
in $\tr\bigl(T_{0} \tr_{\mathrm{L}}(\rho_0) T_{0}^{\dagger}  \bigr)$. The first term of $P_{\mathrm{top}}(t)$ is thus given by the sum over all $k\in{1,2,...,M}$ [see Eq.~(\ref{eq:Atop horizontal})], multiplied by the initial population of the ladder ground state, resulting in
\begin{align}
    \dfrac{1}{\mathcal{Z}_{\mathrm{L}}}\tr\bigl(T_{0} \tr_{\mathrm{L}}(\rho_0) T_{0}^{\dagger}  \bigr) &=\,
    \dfrac{1}{\mathcal{Z}_{\mathrm{L}}}\dfrac{\mathcal{Z}_{\mathrm{H}}-1}{\mathcal{Z}_{\mathrm{H}}\mathcal{Z}_{\mathrm{C}}}
    \sum_{k=1}^{M}\left(\dfrac{1+(\mathcal{Z}_{\mathrm{C}}-1)(\mathcal{Z}_{\mathrm{H}}-1)}{\mathcal{Z}_{\mathrm{H}}\mathcal{Z}_{\mathrm{C}}}\right)^{k-1}\sin^2(gt).
\end{align}
The second part of $P_{\mathrm{top}}(t)$ can be calculated in the same way. Comparing Eq.~(\ref{eq:T1horizontal}) with Eq.~(\ref{eq:Atop horizontal}), one sees that (aside from the oscillating scalar factors) the first term of $T_1$ differs from $T_0$ by replacing $\sigma_{\mathrm{M}_k}^{-}$ in the latter with $\ket{1_\mathrm{C}0_\mathrm{H}}\!\!\bra{1_\mathrm{C}0_\mathrm{H}}_{\mathrm{M}_k}$, and the corresponding factor $\tfrac{\mathcal{Z}_{\mathrm{H}}-1}{\mathcal{Z}_{\mathrm{H}}\mathcal{Z}_{\mathrm{C}}}$ is thus replaced by
\begin{align}
   & \tr\left[ \dfrac{1}{\mathcal{Z}_{\mathrm{H}}\mathcal{Z}_{\mathrm{C}}}\ket{1_\mathrm{C}0_\mathrm{H}}\!\!\bra{1_\mathrm{C}0_\mathrm{H}}_{\mathrm{M}_k}\left(\ket{0}\!\!\bra{0}_{\mathrm{C}_k}+(\mathcal{Z}_{\mathrm{C}}-1)\ket{1}\!\!\bra{1}_{\mathrm{C}_k} \right)\otimes\left(\ket{0}\!\!\bra{0}_{\mathrm{H}_k}+(\mathcal{Z}_{\mathrm{H}}-1)\ket{1}\!\!\bra{1}_{\mathrm{H}_k}\right)\ket{1_\mathrm{C}0_\mathrm{H}}\!\!\bra{1_\mathrm{C}0_\mathrm{H}}_{\mathrm{M}_k}\right]\nonumber\\
    &\ =\tr\left[ \dfrac{1}{\mathcal{Z}_{\mathrm{H}}\mathcal{Z}_{\mathrm{C}}}(\mathcal{Z}_{\mathrm{C}}-1)\ket{1_\mathrm{C}0_\mathrm{H}}\!\!\bra{1_\mathrm{C}0_\mathrm{H}}_{\mathrm{M}_k}  \right]=\dfrac{\mathcal{Z}_{\mathrm{C}}-1}{\mathcal{Z}_{\mathrm{H}}\mathcal{Z}_{\mathrm{C}}} .
\end{align}
The transition operator $T_{1}$ additionally contains the static term $\tilde{\Pi}$, which means that $P_{\mathrm{top}}(t)$ contains the additional term 
\begin{align}
    &\tr\left[\dfrac{1}{\mathcal{Z}_{\mathrm{H}}^M\mathcal{Z}_{\mathrm{C}}^M}\tilde{\Pi}\bigotimes_{k=1}^{M} \left(\ket{0}\!\!\bra{0}_{\mathrm{C}_k}+(\mathcal{Z}_{\mathrm{C}}-1)\ket{1}\!\!\bra{1}_{\mathrm{C}_k} \right)\otimes\left(\ket{0}\!\!\bra{0}_{\mathrm{H}_k}+(\mathcal{Z}_{\mathrm{H}}-1)\ket{1}\!\!\bra{1}_{\mathrm{H}_k}\right) \tilde{\Pi} \right]\nonumber\\
    &=\frac{\left(1+\left(\mathcal{Z}_{\mathrm{C}}-1\right)\left(\mathcal{Z}_{\mathrm{H}}-1\right)\right)^{M}}{\mathcal{Z}_{\mathrm{H}}^M \mathcal{Z}_{\mathrm{C}}^M} +\sum_{k=1}^{M}\frac{\left(1+\left(\mathcal{Z}_{\mathrm{C}}-1\right)\left(\mathcal{Z}_{\mathrm{H}}-1\right)\right)^{k-1}\left(\mathcal{Z}_{\mathrm{H}}-1\right)}{\mathcal{Z}_{\mathrm{H}}^{k-1} \mathcal{Z}_{\mathrm{C}}^{k-1}} .
\end{align}
Thus, the second term of $P_{\mathrm{top}}(t)$ becomes
\begin{align}
   \dfrac{\mathcal{Z}_{\mathrm{L}}-1}{\mathcal{Z}_{\mathrm{L}}}\tr\bigl(T_{1} \tr_{\mathrm{L}}(\rho_0) T_{1}^{\dagger}  \bigr) &=\, \dfrac{\mathcal{Z}_{\mathrm{L}}-1}{\mathcal{Z}_{\mathrm{L}}}\dfrac{\mathcal{Z}_{\mathrm{C}}-1}{\mathcal{Z}_{\mathrm{H}}\mathcal{Z}_{\mathrm{C}}}
    \sum_{k=1}^{M}\left(\dfrac{1+(\mathcal{Z}_{\mathrm{C}}-1)(\mathcal{Z}_{\mathrm{H}}-1)}{\mathcal{Z}_{\mathrm{H}}\mathcal{Z}_{\mathrm{C}}}\right)^{k-1}
    \cos^2(gt)\nonumber\\ &\ +\dfrac{\mathcal{Z}_{\mathrm{L}}-1}{\mathcal{Z}_{\mathrm{L}}}\left( \frac{\left(1+\left(\mathcal{Z}_{\mathrm{C}}-1\right)\left(\mathcal{Z}_{\mathrm{H}}-1\right)\right)^{M}}{\mathcal{Z}_{\mathrm{H}}^M \mathcal{Z}_{\mathrm{C}}^M} +\sum_{k=1}^{M}\frac{\left(1+\left(\mathcal{Z}_{\mathrm{C}}-1\right)\left(\mathcal{Z}_{\mathrm{H}}-1\right)\right)^{k-1}\left(\mathcal{Z}_{\mathrm{H}}-1\right)}{\mathcal{Z}_{\mathrm{H}}^{k-1} \mathcal{Z}_{\mathrm{C}}^{k-1}}\right) ,
\end{align}
and the total top-level probability of the horizontal extension is given by 
\begin{align}
     P_{\mathrm{top}}(t) &=\, \sum_{k=1}^{M}\left(\dfrac{1+(\mathcal{Z}_{\mathrm{C}}-1)(\mathcal{Z}_{\mathrm{H}}-1)}{\mathcal{Z}_{\mathrm{H}}\mathcal{Z}_{\mathrm{C}}}\right)^{k-1}\left(\dfrac{1}{\mathcal{Z}_{\mathrm{L}}}\dfrac{\mathcal{Z}_{\mathrm{H}}-1}{\mathcal{Z}_{\mathrm{H}}\mathcal{Z}_{\mathrm{C}}}
   \sin^2(gt) +  \dfrac{\mathcal{Z}_{\mathrm{L}}-1}{\mathcal{Z}_{\mathrm{L}}}\dfrac{\mathcal{Z}_{\mathrm{C}}-1}{\mathcal{Z}_{\mathrm{H}}\mathcal{Z}_{\mathrm{C}}}
   \cos^2(gt)\right)\nonumber\\
   &\ +\dfrac{\mathcal{Z}_{\mathrm{L}}-1}{\mathcal{Z}_{\mathrm{L}}}\left( \frac{\left(1+\left(\mathcal{Z}_{\mathrm{C}}-1\right)\left(\mathcal{Z}_{\mathrm{H}}-1\right)\right)^{M}}{\mathcal{Z}_{\mathrm{H}}^M \mathcal{Z}_{\mathrm{C}}^M} +\sum_{k=1}^{M}\frac{\left(1+\left(\mathcal{Z}_{\mathrm{C}}-1\right)\left(\mathcal{Z}_{\mathrm{H}}-1\right)\right)^{k-1}\left(\mathcal{Z}_{\mathrm{H}}-1\right)}{\mathcal{Z}_{\mathrm{H}}^{k-1} \mathcal{Z}_{\mathrm{C}}^{k-1}}\right)
\end{align}
Taking the limit $T_\mathrm{C}\rightarrow 0$, the top-level probability becomes
\begin{align}
    P_{\mathrm{top}}(t)= \left( 1 - \dfrac{1}{\mathcal{Z}_{\mathrm{H}}^M} \right) \sin^2(gt).
\end{align}


\section{Details on the vertical extension}\label{appsec:vertical extension details}

Here, we present a detailed derivation of the top-level probability $P_{\mathrm{top}}$ for the vertical extension of the clockwork. That is, we consider $M(d-1)$ two-qubit machines coupled to the $d$-level ladder, such that $M$ machines non-trivially couple to each of the $d-1$ ladder transitions. For the purpose of this derivation, we will consider the general case that both the cold and hot bath have finite temperatures, in particular, $T_{\mathrm{C}}\geq 0$ and $T_{\mathrm{H}}<\infty$, but we will assume that $T_{\mathrm{C}}<T_{\mathrm{H}}$. To label the different machines, we denote the Hilbert space of the $j^\mathrm{th}$ machine coupling to the $i^\mathrm{th}$ ladder transition (\emph{i.e.}, the transition between the ladder levels $\ket{i-1}_{\mathrm{L}}$ and $\ket{i}_{\mathrm{L}}$) by $M^{i}_{j}$, where $i\in\{0,1,\ldots,d-1\}$ and $j\in\{1,2,\ldots,M\}$. Moreover, we denote the Hilbert space of the collection of all machines within the same `column' (see, e.g., the illustration in Fig.~\ref{horvert}), \emph{i.e.} the collection of $j^\mathrm{th}$ machines for all ladder transitions, as $M_{(j)}:=\bigotimes_{i=1}^{d-1}M_j^i$. Following these conventions, we define the fully (horizontally and vertically) extended interaction Hamiltonian as
\begin{align}
    H_{\mathrm{int}}=\sum_{k=1}^{M} \;
    \bigotimes_{i=1}^{k-1}\mathbbm{1}_{\mathrm{M}_{(i)}}\otimes J_{\mathrm{M}_{(k)}\mathrm{L}} \otimes \bigotimes_{j=k+1}^{M} \Pi_{\mathrm{M}_{(j)}} 
    \,=\,
    \sum_{k=1}^{M}H_{k}.
\end{align}
Here, the operator $H_{k}$ acts non-trivially on the joint Hilbert space $\mathrm{M}_{(k)}$ of the $k^\mathrm{th}$ `column' and the ladder via the operator
\begin{align}
    J_{\mathrm{M}_{(k)}\mathrm{L}}:=ig\sum_{n=1}^{d-1}\sqrt{n(d-n)}\bigg(\ketbra{n_{\mathrm{M}_{(k)}},n_\mathrm{L}}{n-1_{\mathrm{M}_{(k)}}, n-1_\mathrm{L}}-H.c.\bigg).
\end{align}
However, the action of $J_{\mathrm{M}_{(k)}\mathrm{L}}$ is conditioned on the states of the machines corresponding to the `columns' $\mathrm{M}_{(k+1)}$, $\mathrm{M}_{(k+2)}$, ..., $\mathrm{M}_{(M)}$ through the projectors
\begin{align}
    \Pi_{\mathrm{M}_{(k)}} := \mathbbm{1}_{\mathrm{M}_{(k)}}-\sum_{n=0}^{N-1} \ketbra{n_{\mathrm{M}_{(k)}}}{n_{\mathrm{M}_{(k)}}}\,=\,
    \mathbbm{1}_{\mathrm{M}_{(k)}}-\bar{\Pi}_{\mathrm{M}_{(k)}},
    \label{eq:all the projectors}
\end{align}
where $\ket{0_{\mathrm{M}_{(k)}}}:=\bigotimes_{j=1}^{d-1}|0_\mathrm{C}1_\mathrm{H}\rangle_{\mathrm{M}_k^j}$, and the states $\ket{n_{\mathrm{M}_{(k)}}}$ for $n=1,\ldots,d-1$ are defined as
\begin{align} 
    \ket{n_{\mathrm{M}_{(k)}}} := \bigotimes_{j=1}^{n}|1_\mathrm{C}0_\mathrm{H}\rangle_{\mathrm{M}_k^j}\bigotimes_{l=n+1}^{d-1}|0_\mathrm{C}1_\mathrm{H}\rangle_{\mathrm{M}_k^l}\,.
\end{align}
The state $\ket{n_{\mathrm{M}_{(k)}}}$ can be considered to be the $n^\mathrm{th}$ excited state of the $k^\mathrm{th}$ vertical group M$_{(k)}$ in the sense that the first $n$ machines $\mathrm{M}_{k}^{j}$ for $j=1,\ldots,n$ are in the `used' state $\ket{1_\mathrm{C}0_\mathrm{H}}_{\mathrm{M}_{k}^{j}}$, whereas the remaining $d-n+1$ machines $\mathrm{M}_{k}^{l}$ for $l=n+1,\ldots,d-1$ are in the `unused' state $\ket{0_\mathrm{C}1_\mathrm{H}}_{\mathrm{M}_{k}^{l}}$. Similarly, the state $\ket{0_{\mathrm{M}_{(k)}}}$ represents the corresponding `ground state'. 

Further note that $J_{\mathrm{M}_{(k)}\mathrm{L}}$ acts as an effective generator of rotations on the states $\ket{n_\mathrm{L},n_{\mathrm{M}_{(k)}}}$ for $n=0,\ldots,d-1$. To be more precise, $J_{\mathrm{M}_{(k)}\mathrm{L}}$ can be considered to be a spin-$j$ representation (for $j=\tfrac{d-1}{2}$) of the generator of rotations around the $y$-axis on the subspace $\mathcal{W}_{(k)}:=\operatorname{span}\bigl( \{ \ket{n_{\mathrm{M}_{(k)}}} \}_{n=0,\ldots,d-1} \bigr) \subset M_{(k)}$ of the Hilbert space $M_{(k)}$ of the $k^\mathrm{th}$ vertical group of machines. 
Let us further denote the orthogonal complement of $\mathcal{W}_{(k)}$ with respect to $M_{(k)}$ by $\mathcal{W}_{(k)}^\bot$, such that $M_{(k)}=\mathcal{W}_{(k)}\oplus \mathcal{W}_{(k)}^\bot$. We then observe that $\ker( J_{\mathrm{M}_{(k)}\mathrm{L}})=\mathcal{W}_{(k)}^\bot \otimes\mathcal{H}_\mathrm{L}$. Since $\Pi_{\mathrm{M}_{(k)}}$ projects onto $\mathcal{W}_{(k)}^\bot$, we see that $H_k$ only has support on the subspace $\operatorname{supp}(H_{k})=\bigotimes_{i=1}^{k-1} M_{(i)} \otimes \mathcal{W}_{(k)} \otimes  \bigotimes_{j=k+1}^{m} \mathcal{W}_{(j)}^\bot \otimes \mathcal{H}_\mathrm{L}$ of the total Hilbert space of the ladder and all machines. Moreover, these subspaces are orthogonal for different values of $k$, \emph{i.e.} 
\begin{align}
    H_{k}\,H_{k'}\,=\,0\ \  \forall \ k\,\neq\,k'\,. 
\end{align}
As a consequence, we have $H_{\mathrm{int}}^{q}=\bigl(\sum_{k=1}^{M}H_{k}\bigr)^{q}=\sum_{k=1}^{M}H_{k}^{q}$, which we can use in the power expansion of $e^{-i H_{\mathrm{int}}t}$, that is, 
\begin{align}
    e^{-i H_{\mathrm{int}}t} &=
    \sum\limits_{q=0}^{\infty}\tfrac{(-it)^{q}}{q!}H_{\mathrm{int}}^{q}
    =
    \mathds{1}\,+\,\sum\limits_{q=1}^{\infty}\tfrac{(-it)^{q}}{q!}H_{\mathrm{int}}^{q}
    =
    \mathds{1}\,+\,\sum\limits_{k=1}^{M}\sum\limits_{q=1}^{\infty}\tfrac{(-it)^{q}}{q!}H_{k}^{q}
    =
    \mathds{1}\,+\,\sum\limits_{k=1}^{M}
    \bigotimes_{i=1}^{k-1}\mathbbm{1}_{\mathrm{M}_{(i)}}\otimes \Bigl(\sum\limits_{q=1}^{\infty}\tfrac{(-it)^{q}}{q!}
    J_{\mathrm{M}_{(k)}\mathrm{L}}^{\,q}\Bigr) 
    \otimes \bigotimes_{j=k+1}^{M} \Pi_{\mathrm{M}_{(j)}},
    \label{eq:new derivation step 1}
\end{align}
where we have isolated the leading order term ($q=0$) in the expansion, and used the fact that the $\mathbbm{1}_{\mathrm{M}_{(i)}}$ and $\Pi_{\mathrm{M}_{(j)}}$ are idempotent. Next, we observe that by definition $J_{\mathrm{M}_{(k)}\mathrm{L}}^{\,q}\bar{\Pi}_{\mathrm{M}_{(k)}}=J_{\mathrm{M}_{(k)}\mathrm{L}}^{\,q}$ for all $q\geq1$. We then define the operator $U_{\mathrm{M}_{(k)}\mathrm{L}}(t):=e^{-i J_{\mathrm{M}_{(k)}\mathrm{L}}t}$ and write
\begin{align}
    U_{\mathrm{M}_{(k)}\mathrm{L}}\bar{\Pi}_{\mathrm{M}_{(k)}}
    =e^{-i J_{\mathrm{M}_{(k)}\mathrm{L}}t}\,
    \bar{\Pi}_{\mathrm{M}_{(k)}}
    =\Bigl(\mathds{1}_{\mathrm{M}_{(k)}\mathrm{L}}
    +\sum_{q=1}^{\infty}\tfrac{(-it)^{q}}{q!} J_{\mathrm{M}_{(k)}\mathrm{L}}^{\,q}\bigr) \,
    \bar{\Pi}_{\mathrm{M}_{(k)}}
    =
    \mathds{1}_{\mathrm{L}}\otimes \bar{\Pi}_{\mathrm{M}_{(k)}}
    +\sum_{q=1}^{\infty}\tfrac{(-it)^{q}}{q!} J_{\mathrm{M}_{(k)}\mathrm{L}}^{\,q}.
    \label{eq:new derivation step 2}
\end{align}
Inserting Eq.~(\ref{eq:new derivation step 2}) into Eq.~(\ref{eq:new derivation step 1}), we obtain
\begin{align}
    e^{-i H_{\mathrm{int}}t} &=\,
    \mathds{1}\ -\ 
    \mathds{1}_{\mathrm{L}}\otimes
    \sum\limits_{k=1}^{M}
    \bigotimes_{i=1}^{k-1}\mathbbm{1}_{\mathrm{M}_{(i)}}\otimes
    \bar{\Pi}_{\mathrm{M}_{(k)}}    
    \otimes \bigotimes_{j=k+1}^{M} \Pi_{\mathrm{M}_{(j)}}
    \ +\ 
    \sum\limits_{k=1}^{M}
    \bigotimes_{i=1}^{k-1}\mathbbm{1}_{\mathrm{M}_{(i)}}\otimes 
    U_{\mathrm{M}_{(k)}\mathrm{L}} \,\bar{\Pi}_{\mathrm{M}_{(k)}}
    \otimes \bigotimes_{j=k+1}^{M} \Pi_{\mathrm{M}_{(j)}}
    \label{eq:new derivation step 3}\\
    &=\,
    \mathds{1}\ -\ \mathds{1}_{\mathrm{L}}\otimes\sum\limits_{k=1}^{M}\tilde{\Pi}\subtiny{0}{-1}{[k]}
    \ +\ \sum\limits_{k=1}^{M}\tilde{U}\subtiny{0}{-1}{[k]}
    \,=\,\bar{\mathds{1}}\ +\ \sum\limits_{k=1}^{M}\tilde{U}\subtiny{0}{-1}{[k]},
    \nonumber
\end{align}
where we have defined $\bar{\mathds{1}}:=\mathds{1}\ -\ \mathds{1}_{\mathrm{L}}\otimes \sum_{k=1}^{M}\tilde{\Pi}\subtiny{0}{-1}{[k]}$. The projectors $\tilde{\Pi}\subtiny{0}{-1}{[k]}$ and operators $\tilde{U}\subtiny{0}{-1}{[k]}$ are defined as
\begin{align}
    \tilde{\Pi}\subtiny{0}{-1}{[k]} &:= \bigotimes_{i=1}^{k-1}\mathbbm{1}_{\mathrm{M}_{(i)}}\otimes
    \bar{\Pi}_{\mathrm{M}_{(k)}}    
    \otimes \bigotimes_{j=k+1}^{M} \Pi_{\mathrm{M}_{(j)}},\\ 
    \tilde{U}\subtiny{0}{-1}{[k]} &:=
    \bigotimes_{i=1}^{k-1}\mathbbm{1}_{\mathrm{M}_{(i)}}\otimes 
    U_{\mathrm{M}_{(k)}\mathrm{L}}\,\bar{\Pi}_{\mathrm{M}_{(k)}}
    \otimes \bigotimes_{j=k+1}^{M} \Pi_{\mathrm{M}_{(j)}},
\end{align}
such that $\tilde{U}\subtiny{0}{-1}{[k]}\tilde{U}\subtiny{0}{-1}{[k']} =0$ and $\tilde{\Pi}\subtiny{0}{-1}{[k]}\tilde{\Pi}\subtiny{0}{-1}{[k']} =0\ \forall\,k\neq k'$, while 
$(\mathds{1}_{\mathrm{L}}\otimes\tilde{\Pi}\subtiny{0}{-1}{[k']}) \tilde{U}\subtiny{0}{-1}{[k]}= \tilde{U}\subtiny{0}{-1}{[k]} (\mathds{1}_{\mathrm{L}}\otimes\tilde{\Pi}\subtiny{0}{-1}{[k']}) = \delta_{kk'}\,\tilde{U}\subtiny{0}{-1}{[k]}$ and $\bar{\mathds{1}}\tilde{U}\subtiny{0}{-1}{[k]}=\tilde{U}\subtiny{0}{-1}{[k]}\bar{\mathds{1}}=0$. 

With this, we are now in a position to provide a compact expression of the top-level probability $P_{\mathrm{top}}(t)$, which takes the form
\begin{align}
    P_{\mathrm{top}}(t) &=\, 
    \tr \bigl[ \ketbra{d-1}{d-1}_\mathrm{L} \,\rho \bigr] 
    \,=\, \tr \bigl[ \,_\mathrm{L\!}\bra{d-1} e^{-i H_\mathrm{int} t} \,\rho_{0}\, e^{i H_\mathrm{int} t} \ket{d-1}_\mathrm{L} \bigr]
    \label{eq:new derivation step 4}\\
    &=\,
    \tr \bigl[ \,_\mathrm{L\!}\bra{d-1} 
    \bigl(\bar{\mathds{1}}+ \sum\limits_{k=1}^{M}\tilde{U}\subtiny{0}{-1}{[k]}\bigr) 
    \,\rho_{0}\, 
    \bigl(\bar{\mathds{1}}+ \sum\limits_{k=1}^{M}\tilde{U}\subtiny{0}{1}{[k]}^{\dagger}\bigr)
    \ket{d-1}_\mathrm{L} \bigr]\nonumber\\
    &=\,
    \tr \bigl[ \,_\mathrm{L\!}\bra{d-1} 
    \bar{\mathds{1}}
    \,\rho_{0}\,
    \bar{\mathds{1}}
    \ket{d-1}_\mathrm{L} \bigr]
    +
    \sum\limits_{k=1}^{M}
    \tr \bigl[ \,_\mathrm{L\!}\bra{d-1} 
    \tilde{U}\subtiny{0}{-1}{[k]}
    \,\rho_{0}\,
    \tilde{U}\subtiny{0}{1}{[k]}^{\dagger}
    \ket{d-1}_\mathrm{L} \bigr],\nonumber
\end{align}
where we have used the assumption that the initial state $\rho_{0}$ is diagonal with respect to the joint eigenbasis of the orthogonal projectors $\tilde{\Pi}\subtiny{0}{-1}{[k]}$, which is the case here because the ladder and all machines qubits are initially thermal with respect to either the cold or hot bath. 

For the first term in $P_{\mathrm{top}}$ we then have
\begin{align}
    \tr \bigl[ \,_\mathrm{L\!}\bra{d-1} 
    \bar{\mathds{1}}
    \,\rho_{0}\,
    \bar{\mathds{1}}
    \ket{d-1}_\mathrm{L} \bigr] &=\,
    \tr \bigl[ \,_\mathrm{L\!}\bra{d-1} 
    \bar{\mathds{1}}
    \,\rho_{0}\,
    \ket{d-1}_\mathrm{L} \bigr]
    \,=\,
   \tr \bigl[ \,_\mathrm{L\!}\bra{d-1} 
    \bigl(\mathds{1}\ -\ \mathds{1}_{\mathrm{L}}\otimes\sum_{k=1}^{M}\tilde{\Pi}\subtiny{0}{-1}{[k]} \bigr)
    \,\rho_{0}\,
    \ket{d-1}_\mathrm{L} \bigr]
    \label{eq:new derivation step 5}\\
    &=\,
    \,_\mathrm{L\!}\bra{d-1}\tau_{\mathrm{L}}(\beta_{\mathrm{C}})\ket{d-1}_{\mathrm{L}}\,
    \Bigl(1\,-\,
    \sum\limits_{k=1}^{M}\tr\bigl[\tilde{\Pi}\subtiny{0}{-1}{[k]}\tr_{\mathrm{L}}(\rho_{0})\bigr]\Bigr).\nonumber
\end{align}
Here, we further have
\begin{align}
    \tr\bigl[\tilde{\Pi}\subtiny{0}{-1}{[k]}\tr_{\mathrm{L}}(\rho_{0})\bigr] &=\,
    \tr\bigl[\bigl(
    \bigotimes_{i=1}^{k-1}\mathbbm{1}_{\mathrm{M}_{(i)}}\otimes
    \bar{\Pi}_{\mathrm{M}_{(k)}}    
    \otimes \bigotimes_{j=k+1}^{M} \Pi_{\mathrm{M}_{(j)}}\bigr) 
    \,\bigotimes_{l=1}^{M}\tau_{\mathrm{M}_{(l)}}\bigr]
    \,=\,
    \tr\bigl[\bar{\Pi}_{\mathrm{M}_{(k)}}\tau_{\mathrm{M}_{(k)}}\bigr]\,
    \prod_{j=k+1}^{M}\tr\bigl[\Pi_{\mathrm{M}_{(j)}}\tau_{\mathrm{M}_{(j)}}\bigr]
    \label{eq:new derivation step 6}\\
    &=\,
    \tr\bigl[\bar{\Pi}_{\mathrm{M}_{(k)}}\tau_{\mathrm{M}_{(k)}}\bigr]\,
    \prod_{j=k+1}^{M}\bigl(1-\tr\bigl[\bar{\Pi}_{\mathrm{M}_{(j)}}\tau_{\mathrm{M}_{(j)}}\bigr]\bigr),
    \nonumber
\end{align}
where we can use Eq.~(\ref{eq:all the projectors}) to calculate
\begin{align}
    \tr\bigl[\bar{\Pi}_{\mathrm{M}_{(j)}}\tau_{\mathrm{M}_{(j)}}\bigr] &=\,
    \sum_{n=0}^{N-1} \bra{n_{\mathrm{M}_{(j)}}}
    \tau_{\mathrm{M}_{(j)}}
    \ket{n_{\mathrm{M}_{(j)}}}
    \,=\,
    \sum_{n=0}^{N-1}
    \prod\limits_{i=1}^{n}
    \bra{1_\mathrm{C}0_\mathrm{H}}\tau_{\mathrm{M}_{j}^{i}}\ket{1_\mathrm{C}0_\mathrm{H}}
    \prod\limits_{l=n+1}^{N-1}
    \bra{0_\mathrm{C}1_\mathrm{H}}\tau_{\mathrm{M}_{j}^{l}}\ket{0_\mathrm{C}1_\mathrm{H}}
    \label{eq:new derivation step 7}\\
    &=\,
    \sum_{n=0}^{N-1}
    \bra{1}\tau_{\mathrm{C}}\ket{1}^{n}
    \bra{0}\tau_{\mathrm{H}}\ket{0}^{n}
    \bra{0}\tau_{\mathrm{C}}\ket{0}^{d-n-1}
    \bra{1}\tau_{\mathrm{H}}\ket{1}^{d-n-1}
    \,=\,
    \frac{1}{\mathcal{Z}_{\mathrm{C}}^{d-1}\mathcal{Z}_{\mathrm{H}}^{d-1}}
    \sum_{n=0}^{d-1}
    (\mathcal{Z}_{\mathrm{C}}-1)^{n}(\mathcal{Z}_{\mathrm{H}}-1)^{d-n-1}
    \nonumber\\
    &=\,
    \frac{(\mathcal{Z}_{\mathrm{H}}-1)^{d}-(\mathcal{Z}_{\mathrm{C}}-1)^{d}}{\mathcal{Z}_{\mathrm{H}}^{d-1}\mathcal{Z}_{\mathrm{C}}^{d-1}(\mathcal{Z}_{\mathrm{H}}-\mathcal{Z}_{\mathrm{C}})}.
    \nonumber
\end{align}
Inserting Eqs.~(\ref{eq:new derivation step 7}) and~(\ref{eq:new derivation step 6}) into Eq.~(\ref{eq:new derivation step 5}) and evaluating the sum over $k$, we obtain
\begin{align}
    \tr \bigl[ \,_\mathrm{L\!}\bra{d-1} 
    \bar{\mathds{1}}
    \,\rho_{0}\,
    \bar{\mathds{1}}
    \ket{d-1}_\mathrm{L} \bigr] &=\,
    \,_\mathrm{L\!}\bra{d-1}\tau_{\mathrm{L}}(\beta_{\mathrm{C}})\ket{d-1}_{\mathrm{L}}\,
    \Bigl(1\,-\,\tfrac{(\mathcal{Z}_{\mathrm{H}}-1)^{d}-(\mathcal{Z}_{\mathrm{C}}-1)^{d}}{\mathcal{Z}_{\mathrm{H}}^{d-1}\mathcal{Z}_{\mathrm{C}}^{d-1}(\mathcal{Z}_{\mathrm{H}}-\mathcal{Z}_{\mathrm{C}})}\Bigr)^{M}.
    \label{eq:ptop first time indep contribution}
\end{align}

Turning to the second term of $P_{\mathrm{top}}$ in Eq.~(\ref{eq:new derivation step 4}), we express the individual terms in the sum over $k$ as
\begin{align}
    \tr \bigl[ \,_\mathrm{L\!}\bra{d-1} 
    \tilde{U}\subtiny{0}{-1}{[k]}
    \,\rho_{0}\,
    \tilde{U}\subtiny{0}{1}{[k]}^{\dagger}
    \ket{d-1}_\mathrm{L} \bigr] &=
    \tr \bigl[ \,_\mathrm{L\!}\bra{d-1}
    \bar{\Pi}_{\mathrm{M}_{(k)}}
    U_{\mathrm{M}_{(k)}\mathrm{L}}\,
    \tau_{\mathrm{L}}\otimes\tau_{\mathrm{M}_{(k)}}\,
    U_{\mathrm{M}_{(k)}\mathrm{L}}^{\dagger}
    \bar{\Pi}_{\mathrm{M}_{(k)}}
    \ket{d-1}_\mathrm{L} \bigr]
    \prod_{j=k+1}^{M}\tr\bigl[\Pi_{\mathrm{M}_{(j)}}\tau_{\mathrm{M}_{(j)}}\bigr].
    \label{eq:new derivation step 8}
\end{align}
Then, we note that we can write
\begin{align}
    \bar{\Pi}_{\mathrm{M}_{(k)}}
    U_{\mathrm{M}_{(k)}\mathrm{L}} &=\,
    \bar{\Pi}_{\mathrm{M}_{(k)}}\otimes\mathds{1}_{\mathrm{L}}\,+\,
    \sum\limits_{q=1}^{\infty}\tfrac{(-it)^{q}}{q!}
    J_{\mathrm{M}_{(k)}\mathrm{L}}^{\,q}
    \,=\,
    \sum_{m,n=0}^{N-1}\ketbra{m_{\mathrm{M}_{(k)}},n_\mathrm{L}}{m_{\mathrm{M}_{(k)}}, n_\mathrm{L}}
    \,+\,
    \sum\limits_{q=1}^{\infty}\tfrac{(-it)^{q}}{q!}
    J_{\mathrm{M}_{(k)}\mathrm{L}}^{\,q}
    \label{eq:new derivation step 9}\\
    &=\,
    \sum_{\substack{m,n=0\\ m\neq n}}^{d-1}\ketbra{m_{\mathrm{M}_{(k)}},n_\mathrm{L}}{m_{\mathrm{M}_{(k)}}, n_\mathrm{L}}\,+\,
    \sum_{n=0}^{d-1}\ketbra{n_{\mathrm{M}_{(k)}},n_\mathrm{L}}{n_{\mathrm{M}_{(k)}}, n_\mathrm{L}}
    \,+\,
    \sum\limits_{q=1}^{\infty}\tfrac{(-it)^{q}}{q!}
    J_{\mathrm{M}_{(k)}\mathrm{L}}^{\,q},\nonumber\\
    &=\,
    \sum_{\substack{m,n=0\\ m\neq n}}^{N-1}\ketbra{m_{\mathrm{M}_{(k)}},n_\mathrm{L}}{m_{\mathrm{M}_{(k)}}, n_\mathrm{L}}\,+\,
    \sum_{n=0}^{d-1}\ketbra{n_{\mathrm{M}_{(k)}},n_\mathrm{L}}{n_{\mathrm{M}_{(k)}}, n_\mathrm{L}}\,e^{-i J_{\mathrm{M}_{(k)}\mathrm{L}}t},\nonumber
\end{align}
where we have separated the terms corresponding to projectors onto the kernel and support of the operator $J_{\mathrm{M}_{(k)}\mathrm{L}}$ in the second step. With this, we can simplify the first factor appearing on the right-hand side of Eq.~(\ref{eq:new derivation step 8}) to
\begin{align}
&\ \tr \bigl[ \,_\mathrm{L\!}\bra{d-1}
    \bar{\Pi}_{\mathrm{M}_{(k)}}
    U_{\mathrm{M}_{(k)}\mathrm{L}}\,
    \tau_{\mathrm{L}}\otimes\tau_{\mathrm{M}_{(k)}}\,
    U_{\mathrm{M}_{(k)}\mathrm{L}}^{\dagger}
    \bar{\Pi}_{\mathrm{M}_{(k)}}
    \ket{d-1}_\mathrm{L} \bigr]
    \,=\,
    \sum\limits_{n=0}^{d-2}
    \bra{n_{\mathrm{M}_{(k)}},d-1_\mathrm{L}}
    \tau_{\mathrm{L}}\otimes\tau_{\mathrm{M}_{(k)}}
    \ket{n_{\mathrm{M}_{(k)}},d-1_\mathrm{L}}
    \label{eq:new derivation step 10}\\
&\hspace*{1cm} +
    \tr \bigl[ \,_\mathrm{L\!}\bra{d-1}
    \Bigl(\sum_{n=0}^{d-1}\ketbra{n_{\mathrm{M}_{(k)}},n_\mathrm{L}}{n_{\mathrm{M}_{(k)}}, n_\mathrm{L}}\,
    e^{-i J_{\mathrm{M}_{(k)}\mathrm{L}}t}\Bigr)
    (\tau_{\mathrm{L}}\otimes\tau_{\mathrm{M}_{(k)}})
    \Bigl(e^{i J_{\mathrm{M}_{(k)}\mathrm{L}}t}\,
    \sum_{n'=0}^{d-1}\ketbra{n'_{\mathrm{M}_{(k)}},n'_\mathrm{L}}{n'_{\mathrm{M}_{(k)}}, n'_\mathrm{L}}\Bigr)
    \ket{d-1}_\mathrm{L} \bigr]    
    \nonumber\\
    &\ \ \ =
    \,_\mathrm{L\!}\bra{d-1}\tau_{\mathrm{L}}(\beta_{\mathrm{C}})\ket{d-1}_{\mathrm{L}}\,
    \sum\limits_{n=0}^{d-2}\bra{n_{\mathrm{M}_{(k)}}}\tau_{\mathrm{M}_{(k)}}\ket{n_{\mathrm{M}_{(k)}}}
    \nonumber\\
    &\hspace*{1cm}+\,
    \sum\limits_{n=0}^{d-1}\bra{n_{\mathrm{M}_{(k)}},n_{\mathrm{L}}}\tau_{\mathrm{M}_{(k)}}\otimes\tau_{\mathrm{L}}\ket{n_{\mathrm{M}_{(k)}},n_{\mathrm{L}}}\ 
    \bigl|\bra{N-1_{\mathrm{M}_{(k)}},d-1_{\mathrm{L}}}
    e^{-i J_{\mathrm{M}_{(k)}\mathrm{L}}t}
    \ket{n_{\mathrm{M}_{(k)}},n_{\mathrm{L}}}\bigr|^{2}\nonumber\\
    &\ \ \ =
    \,_\mathrm{L\!}\bra{d-1}\tau_{\mathrm{L}}(\beta_{\mathrm{C}})\ket{d-1}_{\mathrm{L}}\,
    \sum\limits_{n=0}^{d-2}
    \frac{(\mathcal{Z}_{\mathrm{C}}-1)^{n}(\mathcal{Z}_{\mathrm{H}}-1)^{d-n-1}}{\mathcal{Z}_{\mathrm{C}}^{d-1}\mathcal{Z}_{\mathrm{H}}^{d-1}}\nonumber\\
    &\hspace*{1cm}+\,
    \sum\limits_{n=0}^{d-1}
    \,_\mathrm{L\!}\bra{n}\tau_{\mathrm{L}}(\beta_{\mathrm{C}})\ket{n}_{\mathrm{L}}\,
    \frac{(\mathcal{Z}_{\mathrm{C}}-1)^{n}(\mathcal{Z}_{\mathrm{H}}-1)^{d-n-1}}{\mathcal{Z}_{\mathrm{C}}^{d-1}\mathcal{Z}_{\mathrm{H}}^{d-1}}
    \ 
    \bigl|\bra{d-1_{\mathrm{M}_{(k)}},d-1_{\mathrm{L}}}
    e^{-i J_{\mathrm{M}_{(k)}\mathrm{L}}t}
    \ket{n_{\mathrm{M}_{(k)}},n_{\mathrm{L}}}\bigr|^{2}.\nonumber
\end{align}
Reinserting the first term appearing in the last step of Eq.~(\ref{eq:new derivation step 10}) back into Eq.~(\ref{eq:new derivation step 8}), and evaluating the sum over $k$ in Eq.~(\ref{eq:new derivation step 4}), we obtain another [\emph{i.e.} in addition to that in Eq.~(\ref{eq:ptop first time indep contribution})] time-independent contribution to the top-level probability, given by
\begin{align}
    &
    \,_\mathrm{L\!}\bra{d-1}\tau_{\mathrm{L}}(\beta_{\mathrm{C}})\ket{d-1}_{\mathrm{L}}\,
    \sum\limits_{k=1}^{M}
    \Bigl(1\,-\,\tfrac{(\mathcal{Z}_{\mathrm{H}}-1)^{d}-(\mathcal{Z}_{\mathrm{C}}-1)^{d}}{\mathcal{Z}_{\mathrm{H}}^{d-1}\mathcal{Z}_{\mathrm{C}}^{d-1}(\mathcal{Z}_{\mathrm{H}}-\mathcal{Z}_{\mathrm{C}})}\Bigr)^{M-k}\,
    \sum\limits_{n=0}^{d-2}
    \frac{(\mathcal{Z}_{\mathrm{C}}-1)^{n}(\mathcal{Z}_{\mathrm{H}}-1)^{d-n-1}}{\mathcal{Z}_{\mathrm{C}}^{d-1}\mathcal{Z}_{\mathrm{H}}^{d-1}}
    \label{eq:ptop second time indep contribution}\\
    &\ \ =\,
    \,_\mathrm{L\!}\bra{d-1}\tau_{\mathrm{L}}(\beta_{\mathrm{C}})\ket{d-1}_{\mathrm{L}}\,
    \Bigl(1\,-\,\tfrac{(\mathcal{Z}_{\mathrm{C}}-1)^{d-1}(\mathcal{Z}_{\mathrm{H}}-\mathcal{Z}_{\mathrm{C}})}{(\mathcal{Z}_{\mathrm{H}}-1)^{d}-(\mathcal{Z}_{\mathrm{C}}-1)^{d}}\Bigr)
    \Bigl[1-
    \Bigl(1\,-\,\tfrac{(\mathcal{Z}_{\mathrm{H}}-1)^{d}-(\mathcal{Z}_{\mathrm{C}}-1)^{d}}{\mathcal{Z}_{\mathrm{H}}^{d-1}\mathcal{Z}_{\mathrm{C}}^{d-1}(\mathcal{Z}_{\mathrm{H}}-\mathcal{Z}_{\mathrm{C}})}\Bigr)^{M}\Bigr].
    \nonumber
\end{align}

For the second term appearing in the last step of Eq.~(\ref{eq:new derivation step 10}), we note that, since $J_{\mathrm{M}_{(k)}\mathrm{L}}$ corresponds to the spin-$j$ representation (with $j=\tfrac{d-1}{2}$) of the generator of rotations around the $y$-axis on the subspace spanned by the vectors $\ket{n_{\mathrm{M}_{(k)}},n_{\mathrm{L}}}$ for $n=0,1,\ldots,d-1$, the matrix elements $\bra{d-1_{\mathrm{M}_{(k)}},d-1_{\mathrm{L}}} e^{-i J_{\mathrm{M}_{(k)}\mathrm{L}}t} \ket{n_{\mathrm{M}_{(k)}},n_{\mathrm{L}}}$ coincide with the elements of the Wigner (small) d-matrix $d_{\mu,m}^{j}(\beta):=\bra{j,\mu} e^{-i \beta J_y} \ket{j,m}$ for $\mu=j$, $m=n-j$, and $\beta = 2gt$, see, e.g.,~\cite{MorrisonParker1987} or~\cite{Wigner1959GroupTheory}. In particular, Eq.~(B7) in~\cite[p.~485]{MorrisonParker1987} lets us write
\begin{align}
    \bigl|\bra{d-1_{\mathrm{M}_{(k)}},d-1_{\mathrm{L}}}
    e^{-i J_{\mathrm{M}_{(k)}\mathrm{L}}t}
    \ket{n_{\mathrm{M}_{(k)}},n_{\mathrm{L}}}\bigr|^{2}
    &=\,
    \binom{d-1}{n}\,\cos^{2n}(gt)\,\sin^{2(d-n-1)}(gt)\,.
    \label{eq:time dep part}
\end{align}
The prefactors of these sinusoidal contributions are then obtained by combining the second term in the last step of Eq.~(\ref{eq:new derivation step 10}) with Eq.~(\ref{eq:new derivation step 8}), and evaluating the sum over $k$ in Eq.~(\ref{eq:new derivation step 4}), which yields
\begin{align}
    &\,_\mathrm{L\!}\bra{n}\tau_{\mathrm{L}}(\beta_{\mathrm{C}})\ket{n}_{\mathrm{L}}\,
    \tfrac{(\mathcal{Z}_{\mathrm{C}}-1)^{n}(\mathcal{Z}_{\mathrm{H}}-1)^{d-n-1}}{\mathcal{Z}_{\mathrm{C}}^{d-1}\mathcal{Z}_{\mathrm{H}}^{d-1}}
    \sum\limits_{k=1}^{M}
    \Bigl(1\,-\,\tfrac{(\mathcal{Z}_{\mathrm{H}}-1)^{d}-(\mathcal{Z}_{\mathrm{C}}-1)^{d}}{\mathcal{Z}_{\mathrm{H}}^{d-1}\mathcal{Z}_{\mathrm{C}}^{d-1}(\mathcal{Z}_{\mathrm{H}}-\mathcal{Z}_{\mathrm{C}})}\Bigr)^{M-k}
    \label{eq:time dep contr prefactor}\\
    &\ \ =\,
    \,_\mathrm{L\!}\bra{n}\tau_{\mathrm{L}}(\beta_{\mathrm{C}})\ket{n}_{\mathrm{L}}\,
    \tfrac{(\mathcal{Z}_{\mathrm{C}}-1)^{n}(\mathcal{Z}_{\mathrm{H}}-1)^{d-n-1}
    (\mathcal{Z}_{\mathrm{H}}-\mathcal{Z}_{\mathrm{C}})}{(\mathcal{Z}_{\mathrm{H}}-1)^{d}-(\mathcal{Z}_{\mathrm{C}}-1)^{d}}
    \Bigl[1-
    \Bigl(1\,-\,\tfrac{(\mathcal{Z}_{\mathrm{H}}-1)^{d}-(\mathcal{Z}_{\mathrm{C}}-1)^{d}}{\mathcal{Z}_{\mathrm{H}}^{d-1}\mathcal{Z}_{\mathrm{C}}^{d-1}(\mathcal{Z}_{\mathrm{H}}-\mathcal{Z}_{\mathrm{C}})}\Bigr)^{M}\Bigr].
    \nonumber
\end{align}
Finally, we can collect Eqs.~(\ref{eq:time dep part}) and~(\ref{eq:time dep contr prefactor}), and combine them with the time-independent terms in $P_{\mathrm{top}}$ to arrive at
\begin{align}
    P_{\mathrm{top}}(t) &=\,
    \sum\limits_{n=0}^{d-2}
    \,_\mathrm{L\!}\bra{n}\tau_{\mathrm{L}}(\beta_{\mathrm{C}})\ket{n}_{\mathrm{L}}\,
    (\mathcal{Z}_{\mathrm{C}}-1)^{n}(\mathcal{Z}_{\mathrm{H}}-1)^{d-n-1}\, f(M,d,\beta_{\mathrm{C}},\beta_{\mathrm{H}})\,
    \tbinom{d-1}{n}\,\cos^{2n}(gt)\,\sin^{2(d-n-1)}(gt)\nonumber\\
    &\ \ +\,
    \,_\mathrm{L\!}\bra{d-1}\tau_{\mathrm{L}}(\beta_{\mathrm{C}})\ket{d-1}_{\mathrm{L}}\,\Bigl[1-\bigl(1-\cos^{2(d-1)}(gt)\bigr)(\mathcal{Z}_{\mathrm{C}}-1)^{d-1}
    f(M,d,\beta_{\mathrm{C}},\beta_{\mathrm{H}})\Bigr],
    \label{eq:ptop general}
\end{align}
where the coefficient  $f(M,d,\beta_{\mathrm{C}},\beta_{\mathrm{H}})$ is given by
\begin{align}
    f(M,d,\beta_{\mathrm{C}},\beta_{\mathrm{H}}) &=\,
    \tfrac{\mathcal{Z}_{\mathrm{H}}-\mathcal{Z}_{\mathrm{C}}}{(\mathcal{Z}_{\mathrm{H}}-1)^{d}-(\mathcal{Z}_{\mathrm{C}}-1)^{d}}
    \Bigl[1-
    \Bigl(1\,-\,\tfrac{(\mathcal{Z}_{\mathrm{H}}-1)^{d}-(\mathcal{Z}_{\mathrm{C}}-1)^{d}}{\mathcal{Z}_{\mathrm{H}}^{d-1}\mathcal{Z}_{\mathrm{C}}^{d-1}(\mathcal{Z}_{\mathrm{H}}-\mathcal{Z}_{\mathrm{C}})}\Bigr)^{M}\Bigr].
    \label{eq:f}
\end{align}
The expression in Eq.~(\ref{eq:ptop general}) holds for arbitrary temperatures $T_{\mathrm{C}}$ and $T_{\mathrm{H}}>T_{\mathrm{C}}$, and includes the desired term proportional to $\sin^{2(d-1)(gt)}$ in the sum for $n=0$. In particular, this term is the only term in $P_{\mathrm{top}}(t)$ that remains when taking the limit $T_{\mathrm{C}}\rightarrow0$, in which case $\mathcal{Z}_{\mathrm{C}}\rightarrow1$,  $\,_\mathrm{L\!}\bra{d-1}\tau_{\mathrm{L}}(\beta_{\mathrm{C}})\ket{d-1}_{\mathrm{L}}\rightarrow0$, and $\,_\mathrm{L\!}\bra{0}\tau_{\mathrm{L}}(\beta_{\mathrm{C}})\ket{0}_{\mathrm{L}}=1$, and we have 
\begin{align}
    \lim_{T_{\mathrm{C}}\rightarrow0}\,P_{\mathrm{top}}(t) &=\,
    \Bigl[1-
    \Bigl(1\,-\,\bigl(\tfrac{\mathcal{Z}_{\mathrm{H}}-1}{\mathcal{Z}_{\mathrm{H}}}\bigr)^{d-1}\Bigr)^{M}\Bigr]\,
    \sin^{2(d-1)}(gt),
    \label{eq:ptop limit TC to zero}
\end{align}
as stated in Eq.~(\ref{math:ptop}) of the main text.\\

To see that small deviations from the ideal case where $T_\mathrm{C}=0$ still allow for $P_{\mathrm{top}}(t)$ to be close to the corresponding value of the ideal case, \emph{i.e.} to show the stability of our approach to ATPC, we analyse the behaviour of $P_{\mathrm{top}}(t)$ in the limits $M\rightarrow\infty$ and $d\rightarrow\infty$ at finite temperatures. 
To this end we first inspect Eq.~(\ref{eq:f}), and note that the term that is potentiated by $M$ is smaller than $1$. To see this, we first write
\begin{align}
    \frac{(\mathcal{Z}_{\mathrm{H}}-1)^{d}-(\mathcal{Z}_{\mathrm{C}}-1)^{d}}
    {\mathcal{Z}_{\mathrm{H}}^{d-1}\mathcal{Z}_{\mathrm{C}}^{d-1}(\mathcal{Z}_{\mathrm{H}}-\mathcal{Z}_{\mathrm{C}})} &=\,
    \frac{(\mathcal{Z}_{\mathrm{H}}-1)^{d}-(\mathcal{Z}_{\mathrm{C}}-1)^{d}}
    {\mathcal{Z}_{\mathrm{H}}^{d}\mathcal{Z}_{\mathrm{C}}^{d}}\,
    \frac{\mathcal{Z}_{\mathrm{H}}\mathcal{Z}_{\mathrm{C}}}{\mathcal{Z}_{\mathrm{H}}-\mathcal{Z}_{\mathrm{C}}}\,=\,\frac{x^d-y^d}{x-y},
    \label{eq:final proof}
\end{align}
where we have defined $x:=\tfrac{\mathcal{Z}_{\mathrm{H}}-1}{\mathcal{Z}_{\mathrm{H}}\mathcal{Z}_{\mathrm{C}}}$ and $y:=\tfrac{\mathcal{Z}_{\mathrm{C}}-1}{\mathcal{Z}_{\mathrm{H}}\mathcal{Z}_{\mathrm{C}}}$ with the property $0\leq y<x\leq\tfrac{1}{2}$. The expression on the right-hand side of Eq.~(\ref{eq:final proof}) is smaller or equal than $1$ if $x-x^d\geq y-y^d$, which is the case if $x-x^d$ is monotonically increasing on the interval $[0,\tfrac{1}{2}]$. Inspecting the derivative, we have $\tfrac{\partial}{\partial x}(x-x^d)=1-d\,x^{d-1}\geq0$ since $d\,x^{d-1}\leq d/2^{d-1}\leq1$ for $d\geq2$. Consequently, we have $\tfrac{x^d-y^d}{x-y}\leq 1$ and $\lim_{M\rightarrow\infty}\bigl(1-\tfrac{x^d-y^d}{x-y}\bigr)^{M}=0$. 
Therefore, we see that
\begin{align}
    \lim_{M\rightarrow\infty}f(M,d,\beta_{\mathrm{C}},\beta_{\mathrm{H}})=\dfrac{\mathcal{Z}_\mathrm{H}-\mathcal{Z}_\mathrm{C}}{(\mathcal{Z}_\mathrm{H}-1)^d-(\mathcal{Z}_\mathrm{C}-1)^d}\,.
\end{align}
Since we know that $P_\mathrm{top}(t)$ must lie in $[0,1]$, showing that the first term of Eq.~(\ref{eq:ptop general}) ($n=0$) remains close to $1$ when $M$ and $N$ go to infinity is sufficient to show that our approach is stable with respect to deviations from $T_\mathrm{H}\rightarrow\infty$ and $T_\mathrm{C}\rightarrow 0$, \emph{i.e.}
\begin{align}
&\lim_{M\rightarrow\infty}\lim_{d\rightarrow\infty}
\,
   \,_\mathrm{L\!}\bra{0}\tau_{\mathrm{L}}(\beta_{\mathrm{C}})\ket{0}_{\mathrm{L}}\,
    (\mathcal{Z}_{\mathrm{H}}-1)^{d-1}\, f(M,d,\beta_{\mathrm{C}},\beta_{\mathrm{H}})\,
    \tbinom{d-1}{0}\,\sin^{2(d-1)}(gt)\nonumber \nonumber \\
   &\ =\lim_{d\rightarrow\infty}
    \dfrac{1}{\mathcal{Z}_\mathrm{L}}(\mathcal{Z}_\mathrm{H}-1)^{d-1}\dfrac{\mathcal{Z}_\mathrm{H}-\mathcal{Z}_\mathrm{C}}{(\mathcal{Z}_\mathrm{H}-1)^d-(\mathcal{Z}_\mathrm{C}-1)^d}\sin^{2(d-1)}(gt) =\lim_{d\rightarrow\infty} \dfrac{1}{\mathcal{Z}_\mathrm{L}}\dfrac{\mathcal{Z}_\mathrm{H}-\mathcal{Z}_\mathrm{C}}{\mathcal{Z}_\mathrm{H}-1}\dfrac{1}{1-\left(\frac{\mathcal{Z}_\mathrm{C}-1}{\mathcal{Z}_\mathrm{H}-1}\right)^d}\sin^{2(d-1)}(gt)\nonumber\\
    &\ =\begin{cases}
         \dfrac{\mathcal{Z}_\mathrm{H}-\mathcal{Z}_\mathrm{C}}{\mathcal{Z}_\mathrm{L}(\mathcal{Z}_\mathrm{H}-1)},& \text{if } t=\frac{\pi}{2g}\\
         0,              & \text{otherwise}\label{eq:limit stuff}
     \end{cases}. 
\end{align}
The value of the expression in Eq.~(\ref{eq:limit stuff}) for $t=\tfrac{\pi}{2 g}$ can further be written as
\begin{align}
    \dfrac{\mathcal{Z}_\mathrm{H}-\mathcal{Z}_\mathrm{C}}{\mathcal{Z}_\mathrm{L}(\mathcal{Z}_\mathrm{H}-1)}
    &=\dfrac{1+e^{-\beta_\mathrm{H}\mathrm{E}_\mathrm{H}}-(1+e^{-\beta_\mathrm{C}\mathrm{E}_\mathrm{C}})}{e^{-\beta_\mathrm{H}\mathrm{E}_\mathrm{H}}\sum_{n=0}^{d-1}e^{-n\beta_\mathrm{C}(\mathrm{E}_\mathrm{H}-\mathrm{E}_\mathrm{C})}}
    =\dfrac{1-e^{\beta_\mathrm{H}\mathrm{E}_\mathrm{H}-\beta_\mathrm{C}\mathrm{E}_\mathrm{C}}}{1+\sum_{n=1}^{\infty}e^{-n\beta_\mathrm{C}(\mathrm{E}_\mathrm{H}-\mathrm{E}_\mathrm{C})}} \nonumber \\
    &= \label{eFiniteTempCheck} 1 - e^{-\beta_\mathrm{C}(\mathrm{E}_\mathrm{H}-\mathrm{E}_\mathrm{C})}-e^{-(\beta_\mathrm{C}\mathrm{E}_\mathrm{C}-\beta_\mathrm{H}\mathrm{E}_\mathrm{H})}+e^{-\mathrm{E}_\mathrm{H}(\beta_\mathrm{C}-\beta_\mathrm{H})}. 
\end{align}
\hl{Recalling that $\mathrm{E}_\mathrm{H}>\mathrm{E}_\mathrm{C}$, we can see that the quantity in Eq.}~\eqref{eFiniteTempCheck}\hl{  remains close to $1$ for finite temperatures when $\beta_\mathrm{C}\gg\beta_{\mathrm{H}}$ such that $\beta_\mathrm{H}\mathrm{E}_\mathrm{H}<\beta_\mathrm{C}\mathrm{E}_\mathrm{C}$, and when $k_B T_\mathrm{C}<<(\mathrm{E}_\mathrm{H}-\mathrm{E}_\mathrm{C})$, which are both in keeping with the assumptions made in the main text.}


\section{Tick probability density}\label{appsec:Tick probability}

In this appendix we show how our derivation of the tick probability results in an exponential decay. The derivation should not be understood as a new result, but rather as a reminder and clarification.
Treating the decay of the top level as a random event we can approximate its probability of occurring in the time interval $\Delta t$ by
\begin{align}
    \Delta P = \Gamma \Delta t
\end{align}
where $\Gamma$ is given in terms of probability per unit time. In our case here, we have that $\Gamma$ corresponds to the top-level population times the constant $c$, \emph{i.e.}  $\Gamma(t)=P_{\mathrm{top}}(t) c$. Let us denote the cumulative probability that no decay occurred until time $t$ as $P(0,t)$. We can then approximate the probability that no event occurred until $t+\Delta t$ as the probability of no event happening until $t$ times the probability that no event happens in the interval $\Delta t$, \emph{i.e.}
\begin{align}
    &P(0,t+\Delta t)=P(0,t)(1-\Gamma(t)\Delta t)
\end{align} 
which leads to
\begin{align}
    \dfrac{P(0,t+\Delta t)-P(0,t)}{\Delta t}= -\Gamma(t)P(0,t).
\end{align}
If we further let $\Delta t \rightarrow dt$, we get that
\begin{align}
\dfrac{d P(0,t)}{dt}=-\Gamma(t) P(0,t)=-c P_{\mathrm{top}}(t)P(0,t)
 \end{align}
and consequently
\begin{align} 
 P(0,t)=e^{-c\int_0^t P_{\mathrm{top}}(t')dt'}. 
\end{align}
Given this expression for the cumulative probability that no event occurred until time $t$ we can proceed to calculate the probability density of a decay event occurring between time $t$ and $t+dt$. To do so, we differentiate the cumulative probability of having had a decay at time $t$ with respect to $t$, \emph{i.e.} $\tfrac{d[1-P(0,t)]}{dt}$, which results in
\begin{align}
    P_{\mathrm{tick}}(t)=c P_{\mathrm{top}}(t)e^{-c\int P_{\mathrm{top}}(t')dt'}.
\end{align}


\section{Numerical calculation of accuracy and resolution}\label{appsec:numerical}
\label{Simplifying the Integrals}

In order to execute the numerical calculations of the resolution and the accuracy efficiently we need to simplify the necessary integrals [defined in Eqs.~(\ref{eq:resolution}) and (\ref{eq:accuracy})]. In this appendix we present details on our approach to this problem. For simplicity we showcase the calculations for $T_\mathrm{C}\rightarrow0$ and $T_\mathrm{H}\rightarrow\infty$.\\
Assessing resolution and accuracy for a given set of parameters $d,M,c$ and $g$, breaks down to calculating the first and second moment of the tick distribution, \emph{i.e.}
\begin{align}
\overline{t}=\int_0^\infty t P_{\mathrm{tick}}(t) dt&&
\overline{t^2}=\int_0^\infty t^2 P_{\mathrm{tick}}(t) dt
\end{align}
where
\begin{align}
P_{\mathrm{tick}}(t)= c \left\lbrace 1- \left[ 1- \left(\frac{\mathcal{Z}_\mathrm{H}-1}{\mathcal{Z}_\mathrm{H}}\right)^{d-1}\right]^{M} \right\rbrace \sin^{2(d-1)}(gt) \exp\left[-c \left\lbrace 1- \left[ 1- \left(\frac{\mathcal{Z}_\mathrm{H}-1}{\mathcal{Z}_\mathrm{H}}\right)^{d-1}\right]^{M} \right\rbrace \int_0^t dt'\; \sin^{2(d-1)}(g t')\right].
\end{align}
In order to simplify the cumbersome expressions we will use the following notation. We will denote the $k^\mathrm{th}$ moment, as
\begin{align}
I_k=c \left\lbrace 1- \left[ 1- \left(\frac{\mathcal{Z}_\mathrm{H}-1}{\mathcal{Z}_\mathrm{H}}\right)^{d-1}\right]^{M} \right\rbrace \int_0^\infty dt\; t^k \sin^{2(d-1)}(gt)\exp\left[-c \left\lbrace 1- \left[ 1- \left(\frac{\mathcal{Z}_\mathrm{H}-1}{\mathcal{Z}_\mathrm{H}}\right)^{d-1}\right]^{M} \right\rbrace \int_0^t dt'\; \sin^{2(d-1)}(g t') \right].
\end{align}
The \emph{effective coupling} is defined as
\begin{align}
C_{M}:=c \left\lbrace 1- \left[ 1- \left(\frac{\mathcal{Z}_\mathrm{H}-1}{\mathcal{Z}_\mathrm{H}}\right)^{d-1}\right]^{M} \right\rbrace .
\end{align}
Furthermore, let 
\begin{align}
f(t)=C_{M}\int_0^\infty dt'\; \sin^{2(d-1)}(g t').
\end{align}
This leads to a much simpler form for the different moments,
\begin{align}
I_k=C_{M}\int_0^\infty dt\; t^k \sin^{2d}(g t) e^{-f(t)}.
\end{align} 
We can solve the integral in the term $f(t)$ with a solution introduced by Wiener~\cite{Wiener2000}, 
\begin{align}
\int_0^x dx'\; \sin^{2(d-1)}(x')=
\dfrac{1}{4^{d-1}}\left[\binom{2(d-1)}{(d-1)}x +\sum_{p=1}^{d-1} \dfrac{(-1)^p}{p}\binom{2(d-1)}{d-1-p}\sin(2px) \right],
\end{align}
where $\binom{n}{m}=\tfrac{n!}{m!(n-m)!}$ is the binomial coefficient. Employing the solution we get
\begin{align}
f(t)=\dfrac{C_{M}}{4^{d-1}}\left\{ \binom{2(d-1)}{d-1}t+\dfrac{1}{g}\sum_{p=1}^{d-1} \binom{2(d-1)}{d-1-p}\sin(2 p g t) \right\}
\end{align} 
and thus,
\begin{align}
I_k=C_{M}\int_0^\infty dt\; t^k \sin^{2(d-1)}(gt) \exp\left[-\dfrac{C_{M}}{4^{d-1}}\binom{2(d-1)}{d-1} t\right]\times\exp\left[-\dfrac{C_{M}}{4^{d-1}g}\sum_{p=1}^d \dfrac{(-1)^p}{p}\binom{2(d-1)}{d-1-p}\sin(2pgt)\right].
\end{align}
By introducing a `cycle' counting variable  $q=\lfloor\tfrac{gt}{\pi}\rfloor \in \mathds{d}_0$ and its residue $\Theta_q=qt-q\pi$, $\Theta_q\epsilon[0,\pi)$,  \emph{i.e.} substituting with $t=\tfrac{q\pi+\Theta_q}{g}$, we arrive at
\begin{align}
I_k=\dfrac{C_{M}}{g}\sum_{q=0}^\infty \int_0^\pi d\Theta_q\; \left(\dfrac{q\pi +\Theta_q}{g}\right)^k\sin^{2(d-1)}(\Theta_q)&\times \exp\left[-\dfrac{C_{M}}{4^{d-1} g}\binom{2(d-1)}{d-1}(q\pi+\Theta_q) \right]\\
&\times\exp\left[-\dfrac{C_{M}}{4^{d-1}g}\sum_{p=1}^{d-1}\dfrac{(-1)^p}{p}\binom{2(d-1)}{d-1-p}\sin(2p\Theta_q)\right], 
\end{align}
where we have used that $\sin^{2(d-1)}(x\pm n\pi)=\sin^{2(d-1)}(x)$ for $n\epsilon\mathds{Z}$ as well as $\sin[2(x\pm n\pi)]=\sin(2x)$.\\
\newline
We are only interested in explicitly calculating the cases $k=1$ and $k=2$. Note that, $k$ only appears in the term $\left( \tfrac{q\pi + \Theta_q}{g}\right)^k$ and that $\left( \tfrac{q\pi + \Theta_q}{g}\right)^2=\tfrac{1}{g^2}\left[(q\pi)^2+2q\pi\Theta_q+\Theta_q^2 \right]$, which allows us to define the function
\begin{align}
E(\Theta_q)=\exp\left[ -\dfrac{C_{M}}{4^{d-1}g}\binom{2(d-1)}{d-1}\Theta_q\right] \times \exp\left[ -\dfrac{C_{M}}{4^{d-1}g}\sum_{p=1}^{d-1}\dfrac{(-1)^p}{p}\binom{2(d-1)}{d-1-p}\sin(2p\Theta_q)\right],
\end{align}
such that
\begin{align}
I_k=\dfrac{C_{M}}{g}\sum_{q=0}^\infty \exp\left[-\dfrac{C_{M}}{4^{d-1}g} \binom{2(d-1)}{d-1} q\pi \right] \times \int_0^\pi d\Theta_q {\left(\dfrac{q\pi+\Theta_q}{g}\right)^k\sin^{2(d-1)}(\Theta_q)} E(\Theta_q).
\end{align}
 As a last step we observe that $E(\Theta_q)$ does not depend on $q$ directly, but only through $\Theta_q$, so the only direct dependence on $q$ in the integral comes from the term $\left(\tfrac{q\pi+\Theta_q}{g}\right)^k$. This leads us to define 
\begin{align}
\tilde{I}_j := \int_0^\pi d\Theta_q\; \Theta_q^{j}\sin^{2(d-1)}(\Theta_q)E(\Theta_q),
\end{align}
such that we can write the desired first and second moments of the tick distribution as
\begin{align}
I_1=\dfrac{C_{M}}{g^2}\sum_{q=0}^\infty \exp\left[-{\dfrac{C_{M}}{4^{d-1}g}}\binom{2(d-1)}{d-1}q\pi \right]\left\{q\pi\tilde{I}_0 +\tilde{I}_1 \right\}
\end{align}
and
\begin{align}
I_2=\dfrac{C_{M}}{g^3}\sum_{q=0}^\infty \exp\left[-\dfrac{C_{M}}{4^{d-1}g}\binom{2(d-1)}{d-1}q\pi \right]\left\{(q\pi)^2\tilde{I}_0+2q\pi\tilde{I}_1 +\tilde{I}_2 \right\},
\end{align}
respectively. In this way only $\tilde{I}_j$ needs to be calculated numerically for $j=\{0,1,2\}$, which decreases the effective computational costs enormously.

\section{How the `sharpness' of $P_{\mathrm{top}}(t)$ influences accuracy and resolution}\label{appsec:sharpness}

The aim of this appendix is to give further insight about the behaviour of clocks with changing ladder dimension $d$ as well as changing coupling constant $g$, in particular with respect to Figs.~\ref{fig:AccOverN_nonideal},~\ref{fig:A_R} and~\ref{fig:acc_Ediss}  in Sec.~\ref{sec:Numerical results}.

\begin{figure}[htbp!]
  \centering
    (a)\includegraphics[width=0.45\textwidth,trim={0cm 0mm 0cm 0mm}]{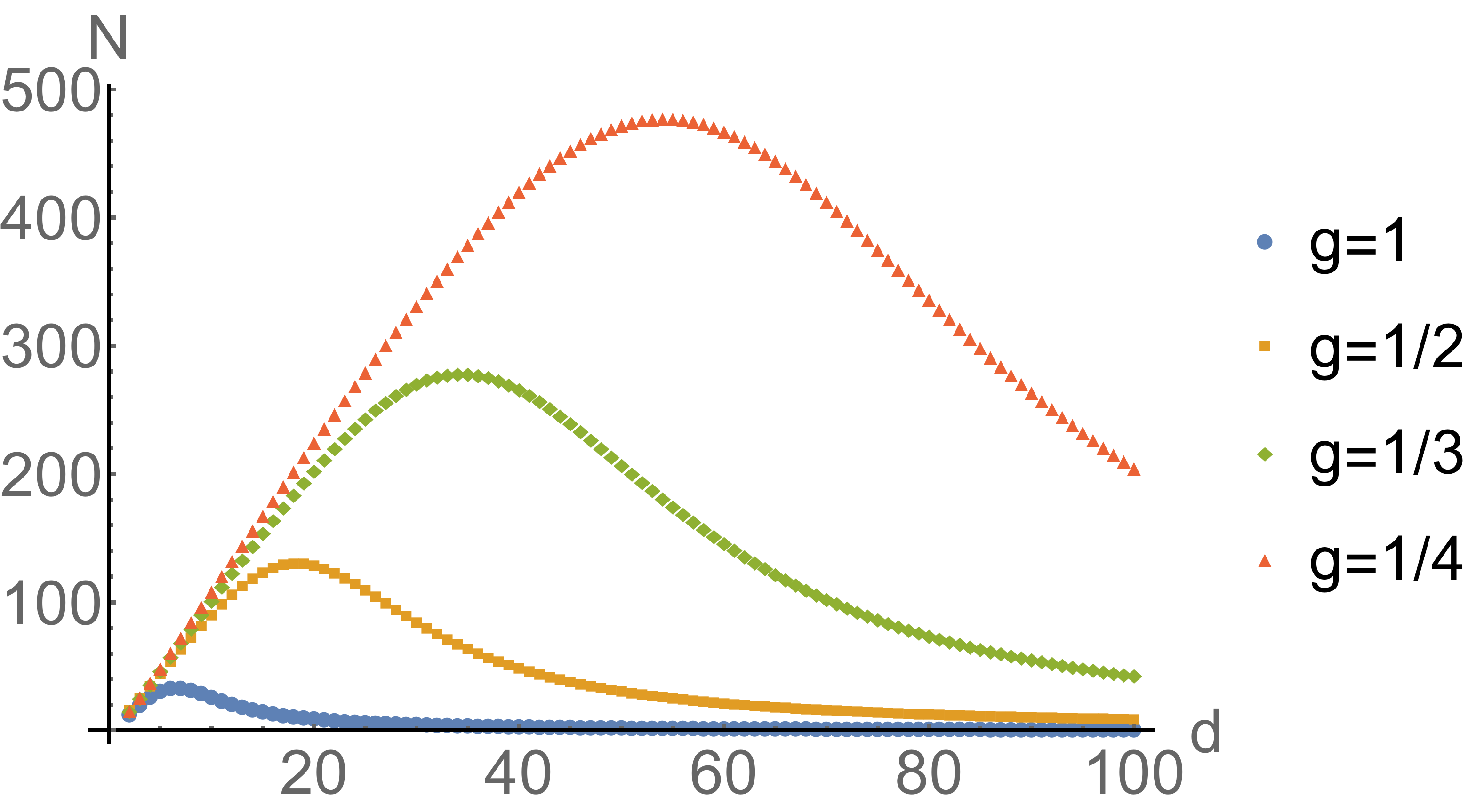}
    (b)\includegraphics[width=0.45\textwidth,trim={0cm 0mm 0cm 0mm}]{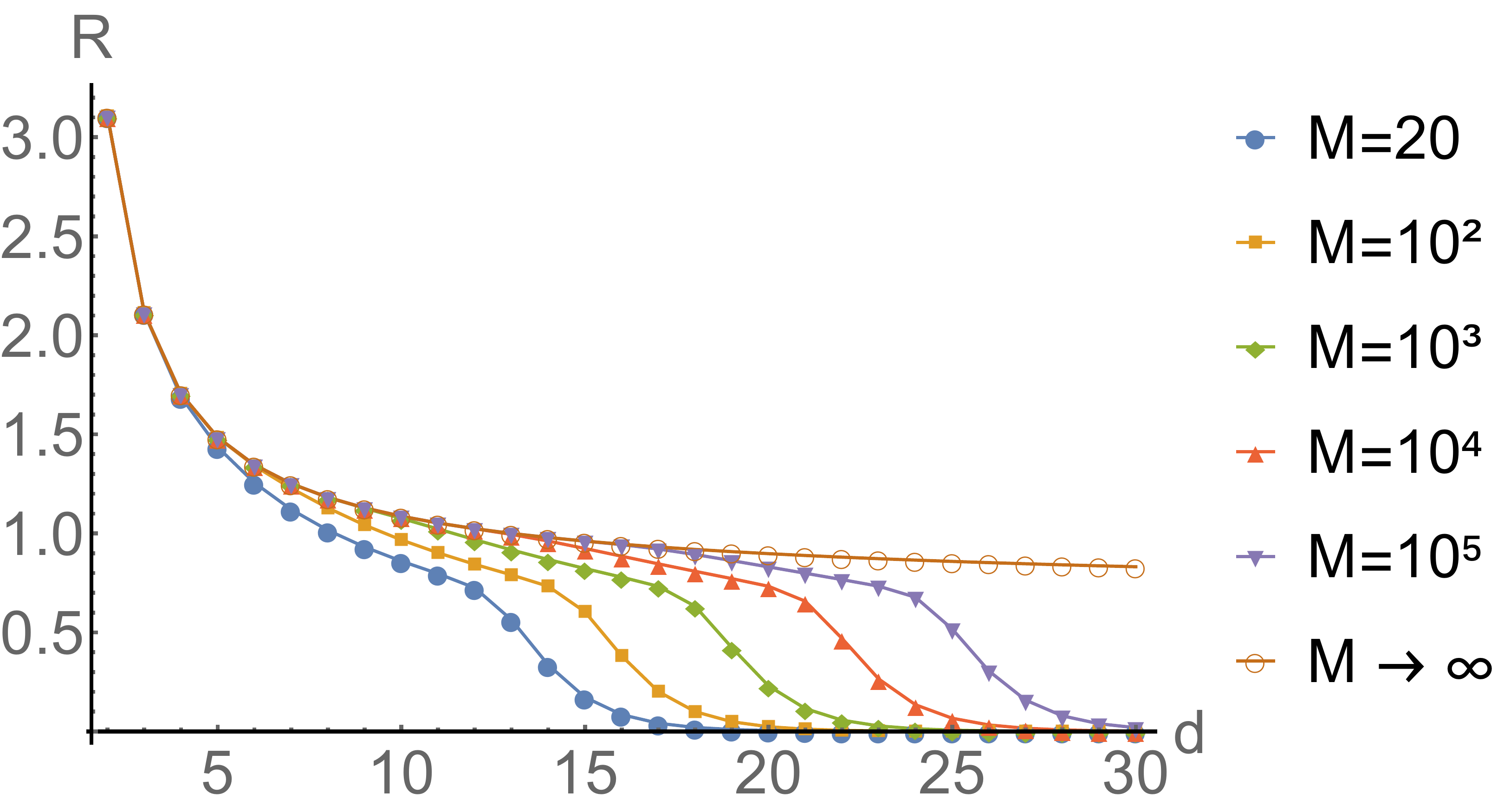}
  \label{fig:sharpness}
\caption{(a) The accuracy is shown as a function of $d$ for different values of $g$ (and fixed $c=10\;\mathrm{s}^{-1}$), where $[g]=\mathrm{E_C}$. The maximally achievable accuracy increases with decreasing $g$. (b) The resolution is shown as a function $d$ (at fixed $g=1\;\mathrm{E_C}$ and $c=1000\;\mathrm{s}^{-1}$). The lines with solid dots show cases of finite $M$. The line with orange circles illustrates the behaviour for the case $M\rightarrow\infty$, which provides an upper bound to the cases with finite $M$. For finite $M$, increasing $d$ reduces the top-level population, eventually becoming so small that the decay event is considerably more likely to skip the first peak of $P_\mathrm{top}(t)$. This leads to an additional reduction in resolution initiating a drop of the resolution eventually approaching $0$ (with $d$).}
\label{fig:acc over N variing g}
\end{figure}

First, let us discuss the relationship between accuracy and `sharpness' of $P_{\mathrm{top}}(t)$. The intuition is that clockworks that are capable of producing a very `sharp' temporal probability distribution should have the potential to give rise to highly accurate clocks, given a suitable irreversible process for the `tick' production. 
Since the maximal amplitude of $P_{\mathrm{top}}(t)$ is given by $ 1- \bigl[ 1- \bigl(\tfrac{\mathcal{Z}_\mathrm{H}-1}{\mathcal{Z}_\mathrm{H}}\bigr)^{d-1}\bigr]^{M} $ (for $T_\mathrm{C}=0$), increasing $M$ leads to an amplitude of $P_\mathrm{top}(t)$ that approaches~$1$ very quickly. Assuming that~$M$ is chosen large enough so that the maximal amplitude is within a desired distance to the value~$1$, the only parameter left that  influences the `sharpness' of the probability distribution is the ladder dimension $d$. We can therefore use~$d$ as a proxy for `sharpness'.  We then proceed by numerically calculating the accuracy in this situation for given values of~$c$ and~$g$. The results are shown in Fig.~\ref{fig:sharpness}~(a) and indicate that the accuracy grows linearly with~$d$. In this regime, the `sharpness' therefore determines the accuracy up to a constant factor. However, one should note that this linear relationship only holds in a regime where the decay process happens fast enough, \emph{i.e.} assuming a sufficiently large value of~$c$ (or small enough value of~$g$). If $P_{\mathrm{top}}(t)$ is too `sharp' compared to the time scale of the decay process increasing the ladder dimension leads to a reduction of the accuracy [as seen in Figs.~\ref{fig:sharpness}(a) and~\ref{fig:AccOverN_nonideal}]. This implies that for a given combination of~$c$ and~$g$ there are certain choices of~$d$ that lead to sub-optimal clocks. 
Considering the resolution as a function of~$d$ [Fig.~\ref{fig:sharpness}(b)] in the limit $M\rightarrow\infty$ we do not observe an optimal configuration. The resolution simply decreases with increasing~$d$ indicating a trade-off relation between accuracy and resolution in the regime where the accuracy increases linearly with~$d$. Thus plotting accuracy over resolution reveals the trade-off relation depicted in Fig.~\ref{fig:A_R}. However, considering finite~$M$ the resolution reaches a point at which it starts dropping to zero quickly. The reason for this can again be found in the amplitude of $P_\mathrm{top}(t)$, which goes to zero for large enough $d$ and fixed $M$. Thus, not only the accuracy (see Fig.~\ref{fig:AccOverN_nonideal}), but also the resolution is bounded from above by the corresponding resolution obtained for $M\rightarrow\infty$. There ~$c$ and~$g$ determine this upper bound.

\begin{figure}[htbp]
  \centering
  \includegraphics[width=0.5\columnwidth]{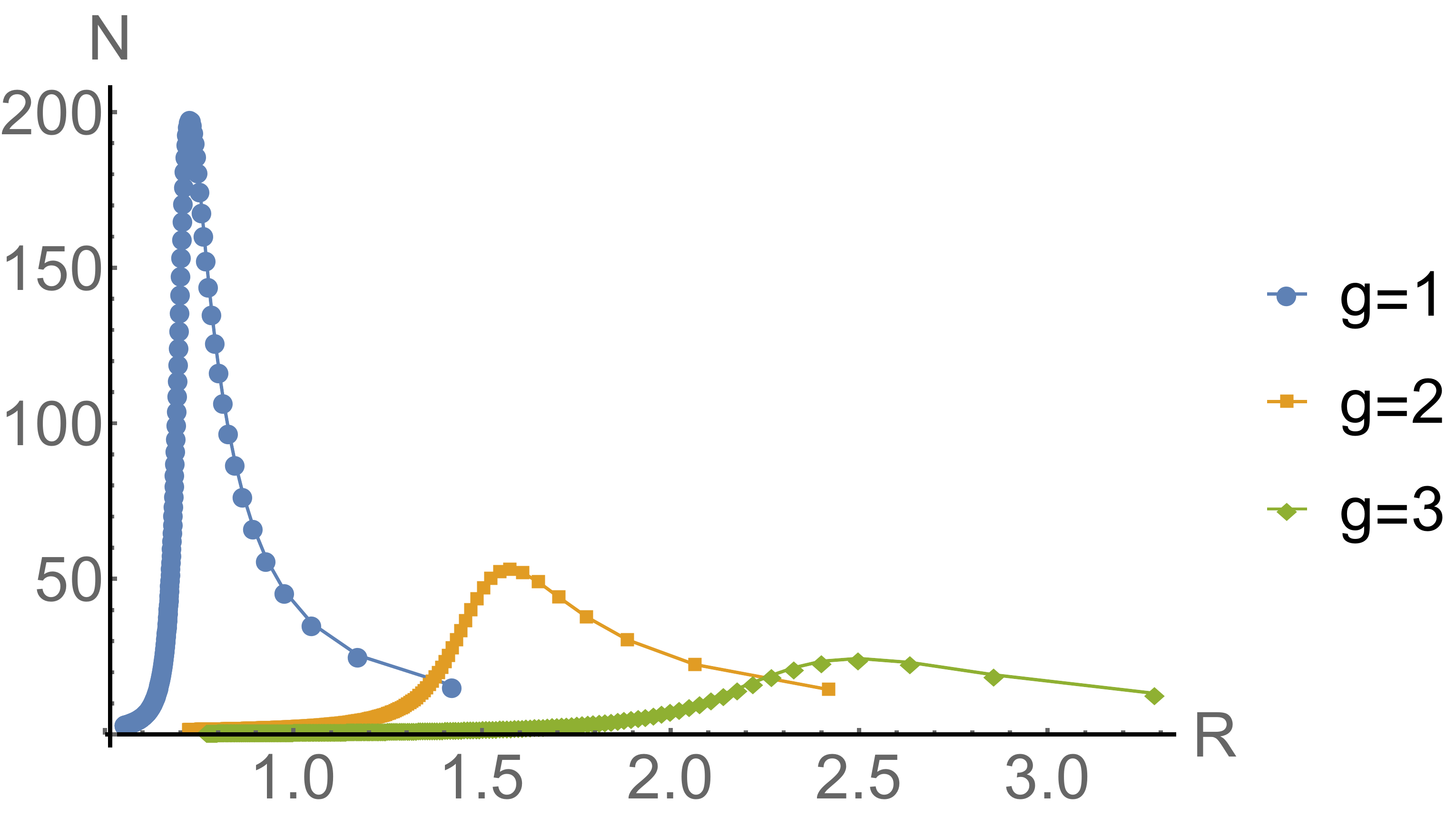}
  \caption{\small{The trade-off between clock accuracy $N$ and resolution $R$ for clockworks of various coupling constants $g$, where $[g]=\mathrm{E_C}$ and $R$ is increased by decreasing $d$. Here we only consider cases of $M\rightarrow\infty$ and $c=25\;\mathrm{s}^{-1}$. Increasing $g$ shifts the peak towards the right, \emph{i.e.} to higher resolutions while decreasing the maximum accuracy. }}
  \label{fig:A_R_g}
\end{figure}

Furthermore, considering only the cases where $M\rightarrow\infty$, Fig.~\ref{fig:A_R_g} shows that increasing~$g$ at fixed~$c$ shifts the point of maximal accuracy to the right, \emph{i.e.} towards higher resolutions. Thus the value of~$g$ determines the lowest resolution at which (optimal) clocks can operate \emph{i.e.} the \emph{maximal cycle-time}. However this increase in resolution comes at the cost of accuracy as increasing $g$ reduces the maximally reachable accuracy. 

\end{document}